\documentclass[12pt]{article}
\usepackage[dvips]{graphicx}

\begin{document}
\begin{center}

\begin{Huge}
\begin{bf}
Vertical Structure Modeling of \\
Saturn's Equatorial Region \\
Using High Spectral Resolution Imaging \\
\end{bf}
\end{Huge}

\vspace{10mm}
{\large\bf T. Temma and N. J. Chanover}\par
Department of Astronomy, MSC 4500 \\
New Mexico State University, P.O.Box 30001 \\ 
Las Cruces, NM 88003-0001 \\
Tel: 505-646-6328 \quad Fax: 505-646-1602 \quad E-mail: temma@nmsu.edu \\
\vspace{5mm}

{\large\bf A. A. Simon-Miller and D. A. Glenar} \\
NASA/Goddard Space Flight Center \\
Greenbelt, MD 20771 \\
\vspace{5mm}

{\large\bf J. J. Hillman} \\
Department of Astronomy, University of Maryland \\
College Park, MD 20742 \\
\vspace{5mm}

{\large\bf D. M. Kuehn}\\
Department of Physics, Pittsburg State University \\
Pittsburg, KS 66762 \\
\vspace{2cm}

Number of pages: 44 \\
\vspace{5mm}
Number of figures: 11 \\
\vspace{5mm}
Number of tables: 7 \\
\vspace{1cm}

Submitted to {\it Icarus} \\
\vspace{5mm}
Submitted July 28, 2003 \\
\vspace{5mm}
Revised --- \\

\end{center}
\clearpage

\vspace{2cm}
{\large\bf Proposed running head:} \\ 
\centerline{Modeling of Saturnian Equatorial Region} \\
\vspace{10mm}

{\large\bf Editorial correspondence to:}
\begin{flushleft}
\begin{large}
Takafumi Temma \\
Department of Astronomy, MSC 4500 \\
New Mexico State University, P.O.Box 30001 \\
Las Cruces, NM 88003-0001 \\
Phone: 505-646-6328 \\
Fax: 505-646-1602 \\
E-mail: temma@nmsu.edu \\
\end{large}
\end{flushleft}
\clearpage

\vspace{2cm}

\centerline{\large\bf Abstract}
A series of narrow-band images of Saturn was acquired using an
Acousto-optic Imaging Spectrometer (AImS) over a large number of 
wavelengths between 500 and 950 nm to perform a detailed study of
Saturn's vertical cloud structure. The Air Force Research Laboratory's
3.67-meter Advanced Electro-Optical System (AEOS) telescope at the Maui
Space Surveillance Complex (MSSC) was used for our observations on 6--11
February 2002. We photometrically calibrated the images with standard
star data to obtain two sets of image cubes of Saturn. The high spectral
resolution ($\Delta \lambda$ = 1.5 -- 5 nm) and wide spectral
coverage of AImS (500 -- 1000 nm) enabled us to sample different
altitudes of the Saturnian equatorial region with higher vertical
resolution than that achievable using conventional narrow-band filters,
and to derive the wavelength dependence of aerosol optical
properties. The theoretical center-limb profiles generated from
radiative transfer computations were fit to the observed center-limb
profiles in the Saturnian equatorial region ($-10^{\circ}$
latitude). Adopting four different cloud structure models with three
different aerosol scattering phase functions, we varied up to nine free
parameters and tried a total of 6000 initial conditions for
optimization to seek the best solution in the vast multi-dimensional
parameter space. Based on the results of the simultaneous fits to five
different profiles around the 890-nm methane band and four profiles
around the 727-nm methane band, we conclude that : 1) a cloud model
having higher aerosol number density in the lower troposphere (0.15 --
1.5 bar) is favorable, 2) the tropospheric cloud extends into the
stratosphere (above 100 mb level), 3) the wavelength dependence of the
upper tropospheric cloud optical thickness indicates a lower limit of
the average aerosol size of roughly 0.7 -- 0.8 $\mu$m, 4) the average
aerosol size of the vertically extended upper tropospheric cloud
increases with depth from about 0.15 $\mu$m in the stratosphere to
between 0.7--0.8 and 1.5 $\mu$m in the troposphere, 5) the aerosol
properties in February 2002 are similar to those seen during the 1990
equatorial disturbance, suggesting a long-term mixing in the upper
atmosphere of Saturn possibly associated with seasonal change.  
\vspace{1cm}

{\large\bf Key Words:} Saturn, atmosphere; atmospheres, structure; photometry

\clearpage

\section{Introduction}

Gas giant planets in the Solar System have intrigued scientists with
their totally different compositions from those of the terrestrial
planets, vigorous atmospheric dynamics and their complex interior
structures. Above all, their compositional similarity to the Sun
inspired scientists to speculate about the formation scenario of the
Solar System. When clouds of ammonia ($NH_3$), ammonium hydro-sulfide
($NH_4SH$) and water ($H_2O$) were theoretically predicted in the
atmospheres of Jovian planets by Weidenschilling and Lewis (1973), this
fundamental result indicated that the vertical cloud structure in a
Jovian planet can tell us about its bulk composition, yielding important
clues about primordial solar system chemistry. Since then, observers
have been motivated to analyze images and spectra of giant planets to
confirm the existence of the predicted cloud decks.    

In the late 1970's, the Pioneer 11 fly-by provided a wealth of
 information about the aerosols in the upper atmosphere of Saturn
 (Tomasko {\it et al.} 1980, Tomasko and Doose 1984). The observations
 with the on-board spectro-polarimeter covered a wide range of phase
 angle, which can not be sampled by ground-based observations, and first
 enabled scientists to characterize the scattering properties of the
 Saturnian aerosols. Subsequently, observational results of ground-based  
 telescopes, the Voyager 2 space probe and the Hubble Space Telescope (HST)
 (West 1983, West {\it et al.} 1983, Karkoschka and Tomasko 1993)
 inferred the aerosol distributions in the Saturnian stratosphere and 
 upper troposphere, assuming diffuse aerosol layers. Karkoschka and
 Tomasko (1992) and Acarreta and S\'anchez-Lavega (1999) placed an
 infinite cloud at the bottom of their model atmospheres to 
 analyze their observational data. Ortiz {\it et al.} (1996) adopted a
 more complicated cloud model similar to those for Jupiter. They examined
 the opacities and altitudes of separated cloud layers. The validity of
 their approach was later corroborated by Stam {\it et al.} (2001), who
 argued for the existence of separated cloud decks based on their
 inversion analysis of near-infrared spectra of Saturn.      

In this paper we use vertical cloud structure models with separated
 cloud decks. These models can also allow for extended diffuse
 clouds. The use of an acousto-optic tunable filter (AOTF) imaging
 camera combined with tip/tilt image compensation resulted in data sets
 superior to previous ones owing to the narrower passbands, finer
 spectral sampling frequency and high spatial resolution. These
 advantages translate into higher vertical resolution for sampling the
 Saturnian atmosphere than previous investigations.    

The main objective of this paper is to examine the Saturnian cloud
structure and aerosol properties, taking full advantage of the unique
features of our observations and our modeling approach. We examine the
vertical aerosol distribution, scattering phase function and the
wavelength dependence of optical properties of aerosols in Saturn's
equatorial region. A comparison between equatorial latitudes and more
southern latitudes will be made in a forthcoming publication. 
        
 We describe our observations and instrument in Section 2. Section 3
 explains the data reduction process. A description of the center-limb
 analysis is presented in Section 4. Section 5 presents the details of
 our modeling approach. After a discussion of the modeling results in
 Section 6, we draw conclusions in Section 7.

\section{Observation, Instrument and Data Set Description}

 Saturn was observed on the nights of 6 -- 11 February 2002 with the
3.67-meter Advanced Electro-Optical System (AEOS) telescope at the Maui
Space Surveillance Complex (MSSC). The tip-tilt correction was on, but
full adaptive optics correction was not used since our target was too
extended. The optical system and the Acousto-optic Imaging Spectrometer
(AImS) were set in a Coud\'e room.  

 AImS was built at NASA's Goddard Space Flight Center as a prototype 
for a Mars lander mission under NASA's Mars Instrument Development
program. This instrument passes the incoming light through a
birefringent crystal called an acousto-optic tunable filter (AOTF), which
operates at wavelengths between 500 and 1000 nm. Monochromatic light is 
diffracted from the crystal in the form of two orthogonally polarized
components, one of which is used for science imaging, and the wavelength
of this light is electronically chosen by applying an acoustic wave of
known frequency into the crystal. The actual spectral bandpass shape
approximates a $sinc^2$ function, and the full-width-half-maximum (FWHM)
of the central lobe is roughly constant in wavenumber units ($\Delta \nu
\sim 40 cm^{-1}$). In wavelength units, the measured FWHM was about 1.5
nm near 500 nm wavelength, and 5 nm around 1000 nm wavelength. More
complete description of AOTF camera operation can be found in Glenar
{\it et al.} (1994,1997), Georgiev {\it et al.} (2002) and Chanover {\it
et al.} (2003). When applied to observations within absorption bands,
this high spectral resolution allows us to sample different atmospheric
levels with a finer vertical resolution than that achieved using
multi-layer or circular variable filters. Using seven separate software
scripts to control AImS automatically, we obtained one set of roughly
160 images of Saturn between 500 and 950 nm on each night of February 7
and 8, 2002 (Fig. 1). A data set of a standard star, HR 2421, was
acquired on each night of February 8 and 11 for the purpose of airmass
correction and comparison with a solar analog star. An image set of a
solar analog star, HR 996 (HD 20630) (Hardorp 1980), was obtained to
calibrate the Saturnian disk intensity in terms of the incident solar
flux on February 11. Exposure times for Saturn were 20 sec in blue
continuum (500--600 nm), 40 sec around weak methane absorption bands and
methane pseudo-continuum (near 619, 727 and 920 nm), and 60 sec near the
890-nm methane absorption band center. These integration times are
longer than those for conventional Saturnian observations because of the
very narrow passband and the polarization selectivity of AImS.    
 
\begin{figure}
\begin{flushleft}
\scalebox{.9}{\includegraphics{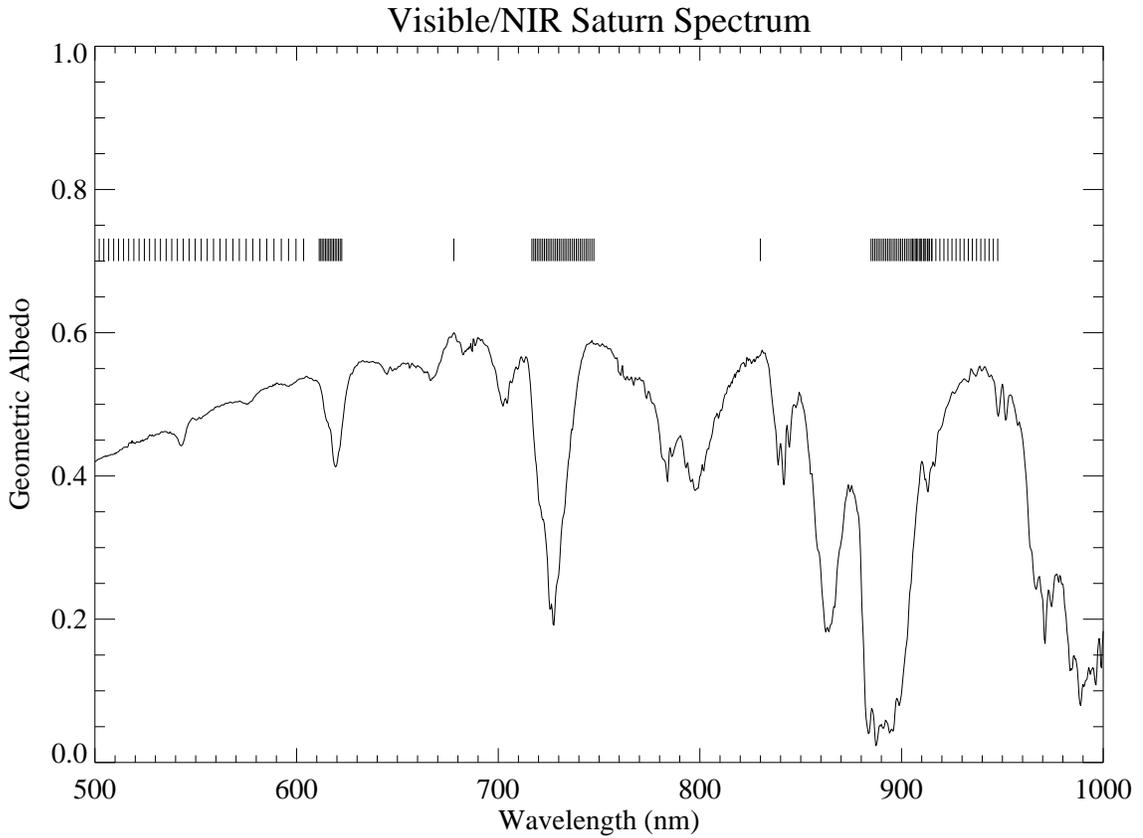}}
\caption{The geometric albedo spectrum of Saturn (Karkoschka 1994) and
 wavelengths at which the Saturnian and stellar images were taken in our
 observations (indicated by short vertical lines). A total of 158 single
 wavelength images were acquired using 7 separate automated sequences of
 blue (500--603 nm, 36 images), green(611--636 nm, 31 images), 678 nm
 continuum (678 nm, 1 image), red (717--747 nm, 34 images), 830 nm
 continuum (830 nm, 1 image), methane band (885--915 nm, 33 images), and
 methane pseudo-continuum (905--948 nm, 22 images).}   
\end{flushleft}
\end{figure}

The spatial resolution of the images was determined from observations of
 a double star, HR 804. We obtained a plate scale of 0.082 $\pm$ 0.002
 arcsec $pix^{-1}$. This corresponds to a horizontal scale of about 540
 km $pix^{-1}$, or $0.5^{\circ}$$pix^{-1}$ of latitude or longitude at
 the sub-earth point. Since we used a guest instrument, the AEOS
 facility de-rotator was not available in the optical axis and therefore
 the field of view slowly rotated on the detector. The smearing effect
 caused by this image rotation is discussed quantitatively in the
 following section.      

At the time of our observations, the solar phase angle was
 $5.9^{\circ}$. The seeing scale was approximately 0.5--0.6 arcsec
 throughout two nights of our observations of Saturn with tip/tilt image
 correction on.

\section{Data Analysis}

\subsection{Image Processing and photometric calibration}

For our modeling analysis, we used the image set of Saturn taken on
February 8, 2002 because of its slightly better seeing ($\sim$ 0.5
arcsec). First, scattered light, bias and dark currents were subtracted
from all the raw images. Second, each image was flat-fielded to remove
pixel-to-pixel sensitivity variations using lamp flat-field
frames. We then applied an airmass correction to the image
intensities, using the data of our secondary standard star, HR 2421. As
described in Appendix A, we photometrically calibrated Saturn by 
comparing it with HR 2421 and subsequently comparing HR 2421 with our
primary standard, HR 996. The calibrated intensity in each pixel is
expressed in terms of $I/F$, which is basically defined as the ratio of
the outgoing intensity $I$ and the incident solar flux $\pi F$. This
process resulted in uncertainties of about 5\% in the calibrated
intensities. Fig. 2 shows the processed images at wavelengths
corresponding to a weak absorption band (727.6 nm), continuum (747.4 nm) 
and strong absorption band (891.3 nm).    

The atmospheric seeing resulted in averaging the surface intensities over
approximately $6\times6$ pixels on the detector. Since we are examining
average center-limb behaviors at different latitudes, we did not find
any need for deconvolution of our images. The influence of the
rotational smearing was estimated by measuring the rotation rate of the
Saturnian polar axis calculated from different images. Fortunately, this
turned out to be only a few pixels (0.1 -- 0.2 arcsec) for each
exposure, and is negligible compared with the seeing.        

We also examined the effect of the scattered light by the Saturnian
rings. We assumed isotropic scattering for simplicity and created a ring
model that contains the two main ring components, the A and B rings. The
reflections from the C ring and other fainter rings were ignored because
of their small opacity and low albedos. The albedos and opacities of the
rings were taken from the Pioneer and Voyager data in Esposito {\it et
al.} (1984). Using those parameters, the contribution to the 
illumination on the Saturnian atmosphere from the two rings was
numerically calculated. The effect of the reflection from the rings was
estimated to be much smaller ($\sim 1\%$) than our total photometry
error ($\sim 5\%$) due to the strong backscattering nature of the
ring particles reported in Esposito {\it et al.} (1984). Therefore, we
neglect this reflection from the rings.          

There was also a concern about the observational bias produced by the
polarization selection associated with the nature of our
instrument. There have been many reports on the degree of polarization of
the Saturnian rings and disk. According to Santer and Dollfus (1981), the
degree of linear polarization around the Saturnian equator is less than
1\% at small phase angles ($\leq 6^{\circ}$) in green light. Dollfus
(1996) observed the polarization of Saturn's rings and disk to show that
the degree of polarization over the disk is less than 1\% in yellow
light. Dulgach {\it et al.} (1983) also stated that the degree of
polarization was less than 1\% at the Saturnian disk center when the
phase angle was less than $\sim 6^{\circ}$ in red light. Consequently,
we assume that the photometric error incurred by the polarization
selection is negligibly small in our analysis.          

\begin{figure}
\begin{center}
\scalebox{.8}{\includegraphics{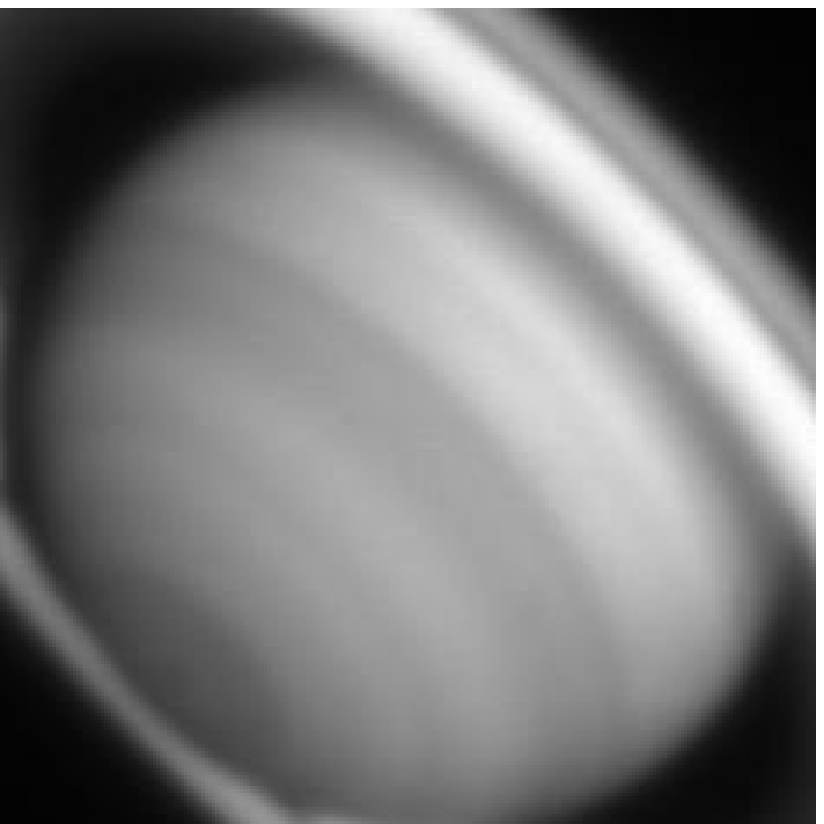}}
\scalebox{.8}{\includegraphics{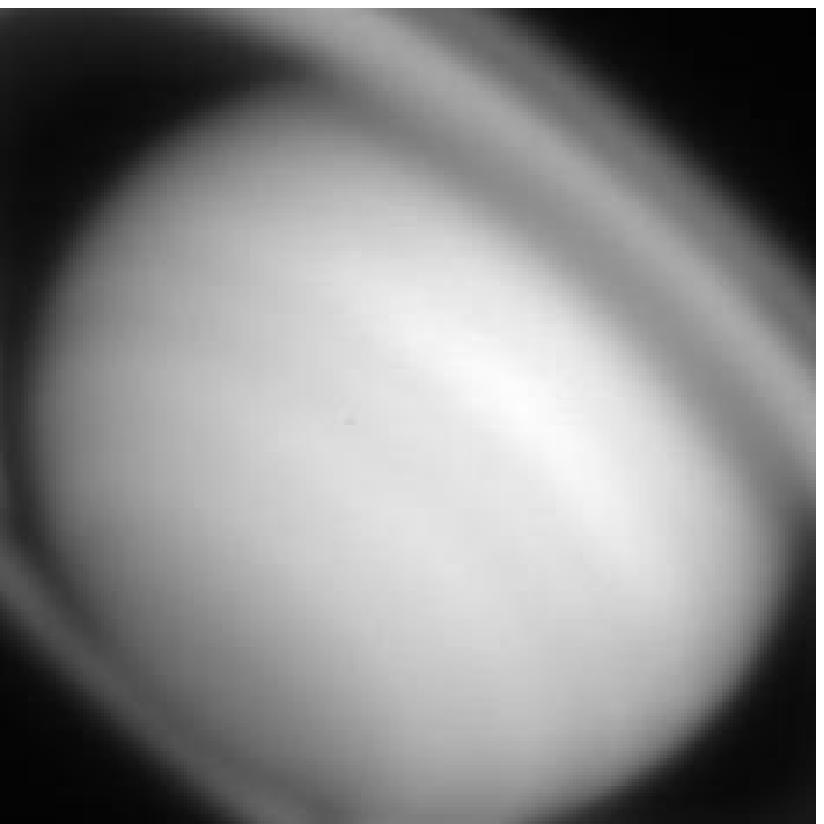}}
\scalebox{.8}{\includegraphics{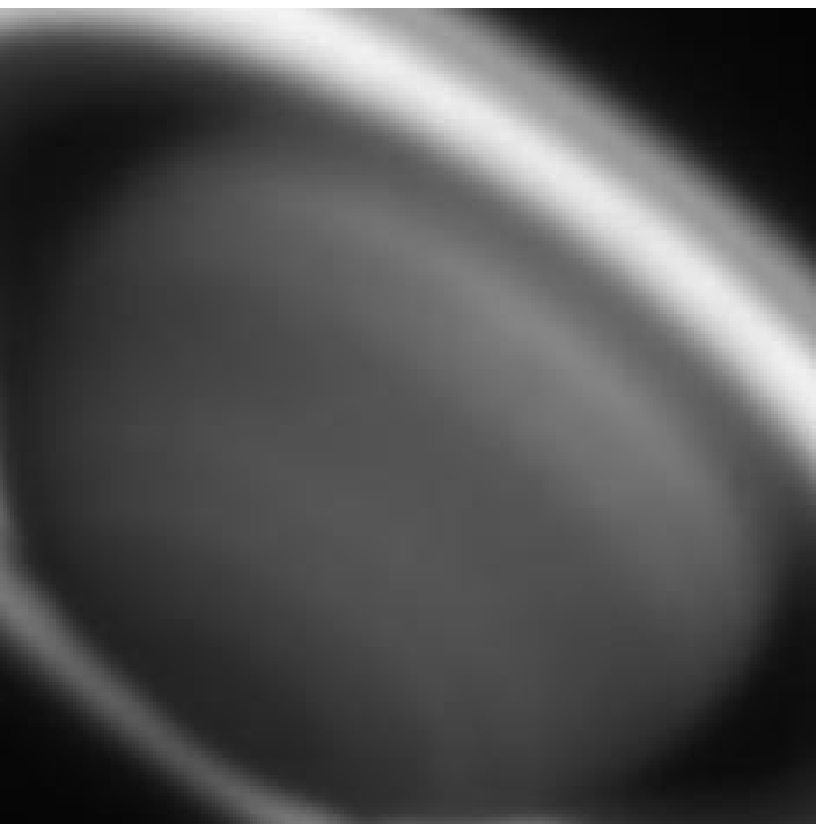}}
\caption{Saturn images taken with AImS at three different wavelengths :
 727.6 nm (top left), 747.4 nm (top right) and 891.3 nm (bottom
 center). Saturnian north is to the upper right, and west
 longitudes increase towards the left side of each image. The 727.6- and
 747.4-nm images were obtained about UT 7:00 February 8, 2002, and the
 891.3-nm image was taken around 30 minutes later.} 
\end{center}
\end{figure}

\subsection{Geometric calibration}

We fit the limb and ring profiles of each image to define the surface
coordinates on Saturn. This process maps a latitude-longitude grid onto the
x-y images of Saturn by using calculated ephemerides, plate scale,
ellipsoidal planetary model and polar orientation information. A
least-squares limb-fitting procedure was used to calculate the exact
position of Saturn's center on each image. The formal error in the 
limb-fitting algorithm yielded the planet center location to within an
accuracy of about $\pm$ 0.5 pixel. Since the limb model does not account
for changes in the atmospheric seeing, the true accuracy of the planet
center determination is slightly worse ($\pm$ 1 pixel). This adds
another $2$\% of error to the model intensity computation. As a result,
we adopt an uncertainty of 7\%, which consists of 5\% photometric
uncertainty and 2\% image navigation error, in the comparison between
the observed intensities and model computations. 
         
In this paper, we chose the equatorial region (planetographic
 latitude of $-10^{\circ}$) for detailed modeling. The relative
 longitude of the sampled region ranges from -$60^{\circ}$ to
 +$60^{\circ}$ from the central meridian, accounting
 for approximately 16 arcsec spanning roughly 200 pixels on
 our detector. We selected 60 data points from the region to obtain
 limb-darkening profiles at selected wavelengths. The cosines of solar
 and Earth zenith angle at each pixel ($\mu_0$ and $\mu$, respectively)
 are always larger than 0.4. This guarantees that the plane-parallel
 approximation is valid in the model intensity computation.

\section{Principle of center-limb analysis and Wavelength selection}

Center-limb profile analysis is a powerful tool to study the vertical
 structure of an atmosphere. The variation of scattering geometry along
 the profile enables us to sense the optical thickness and altitude of
 aerosol layers. Center-limb profiles at methane band centers contain
 aerosol information of only the highest region of the atmosphere
 because the strong absorption hinders deep penetration of the light
 into the atmosphere. Profiles in weaker absorption bands and continua
 contain information of the deeper atmosphere. Thus, examining
 center-limb profiles at different wavelengths of different absorption
 strength is equivalent to 'peeling off' atmospheric layers one by one
 to determine the vertical cloud structure. From the wide spectral range
 of our entire data sets, we selected two clusters of wavelengths
 around the 727- and 890-nm methane bands. The reason is two-fold : 1)
 this selection minimizes the effect of the gaseous Rayleigh scattering,
 which becomes significant at shorter wavelengths, 2) the 619-nm methane
 absorption is too weak to create any physically meaningful contrast
 between the band center and nearby continuum under our total
 calibration error.        

The variation in sampled depths as a function of wavelength is nicely
visualized with contribution functions. The peak position of a
contribution function illustrates where a major fraction of scattering
occurs in a cloudless atmosphere at a certain wavelength. In other
words, the peak location shows the altitude to which the center-limb
profile of a chosen wavelength can have high sensitivity to the
existence of scattering aerosols, provided there is no large aerosol
opacity at higher altitudes. Therefore, when there are a number of
observations at different wavelengths, narrowly-spaced contribution
function peaks signify high vertical sampling resolution. However, on
account of observational errors and variable image quality, it is
difficult to select a set of wavelengths whose contribution function
peaks are narrowly and evenly spaced. In our case, given our 7\%
uncertainty in the calibrated intensities, it was necessary to select
the wavelengths where the observed intensities at a certain position on
the Saturnian disk differ from one another by more than that
amount. Bearing this argument in mind, we selected four wavelengths
(726.5, 732.1, 735.8 and 747.4 nm) around the 727-nm methane band, and
five wavelengths (891.3, 899.6, 906.2, 914.9 and 939.3 nm) around the
890-nm band to maximize the vertical resolution under our calibration
uncertainty. These nine selected wavelengths span from the band centers
to the adjacent continua as shown in Fig. 3. Hereafter, we refer to the
former group of data as the '727-nm data set', and the latter as the
'890-nm data set'.    
\vspace{5mm}

\begin{figure}
\begin{center}
\includegraphics{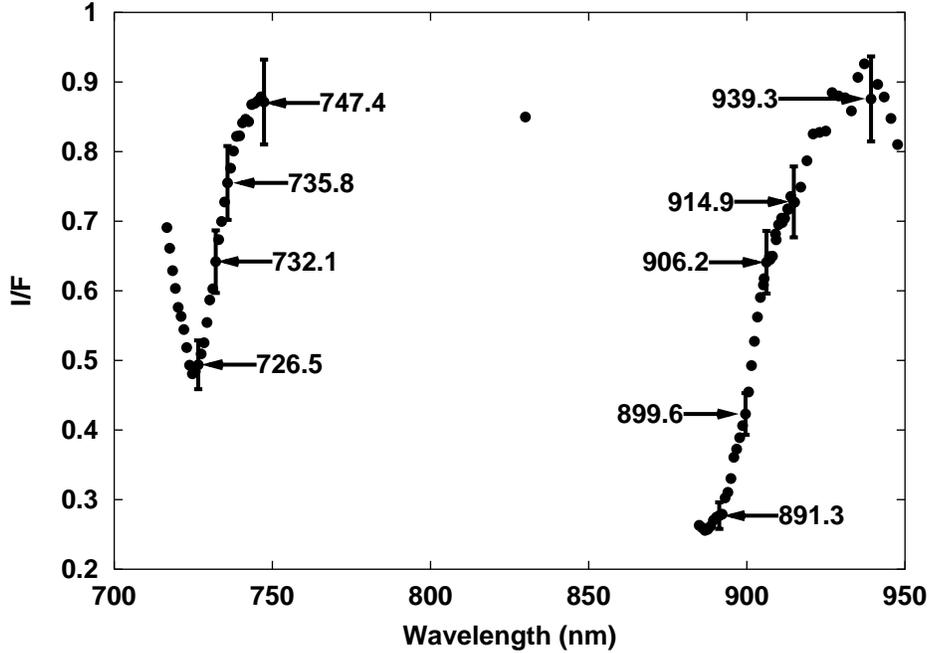}
\caption{The spectral positions of our images between 700 and 950
 nm. The data points were taken from the equatorial region at
 $-10^{\circ}$ planetographic latitude at the central meridian. The
 error bars represent 7\% calibration error at the nine selected
 wavelengths in our modeling analysis.} 
\end{center}
\end{figure}

The locations of the contribution function peaks at our selected
wavelengths are displayed in Fig. 4. They were computed following
the formulation of Banfield {\it et al.} (1996) and using the
temperature profile of Lindal {\it et al.} (1985), under the assumption
of a single-scattering cloudless atmosphere and vertical solar
incidence. Though the single-scattering assumption may be inappropriate
in weak methane bands and continuum, the obtained peak positions still
show the approximate pressure levels at which the major fraction of
scattering occurs. In the absence of aerosols, our 727- and 890-nm data
sets can probe down to roughly 5- and 10-bar depths, respectively. 

\begin{figure}
\begin{center}
\includegraphics{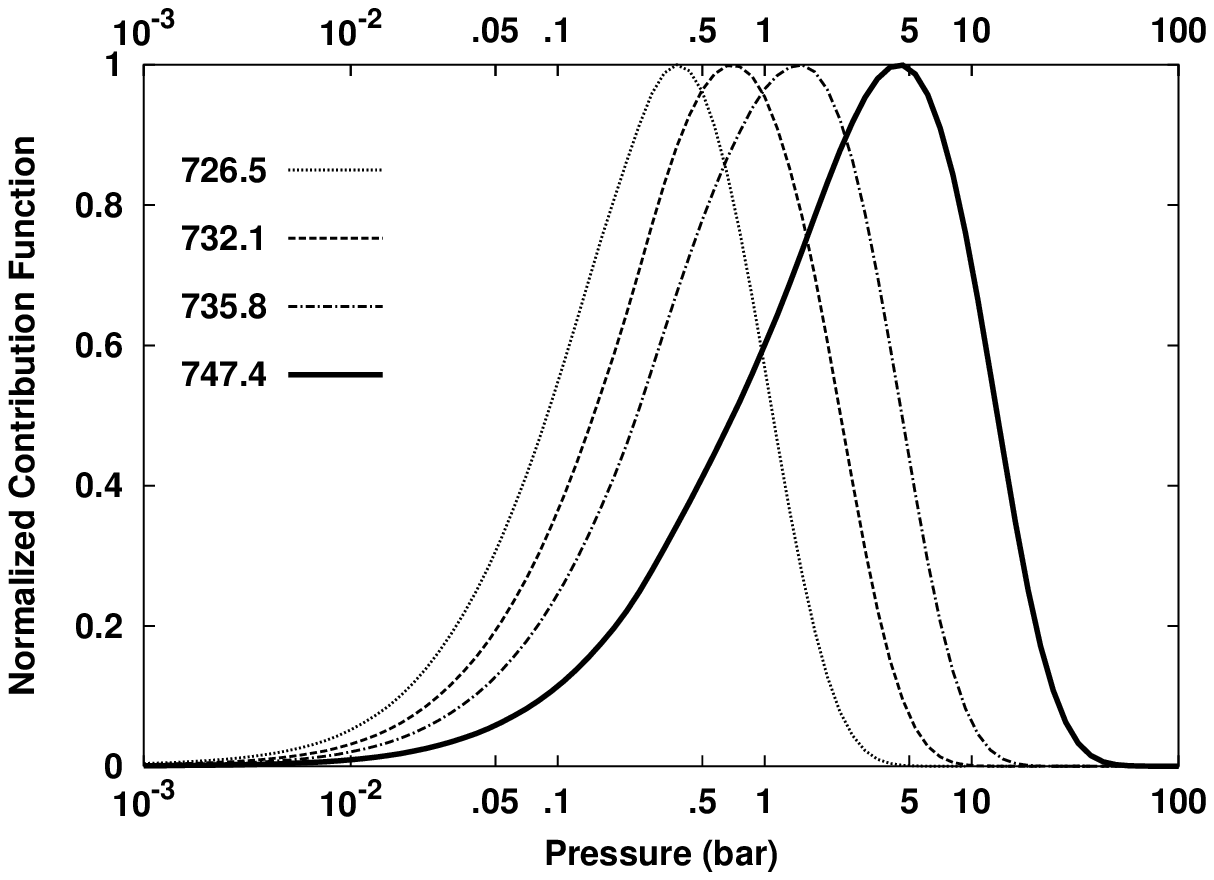}
\includegraphics{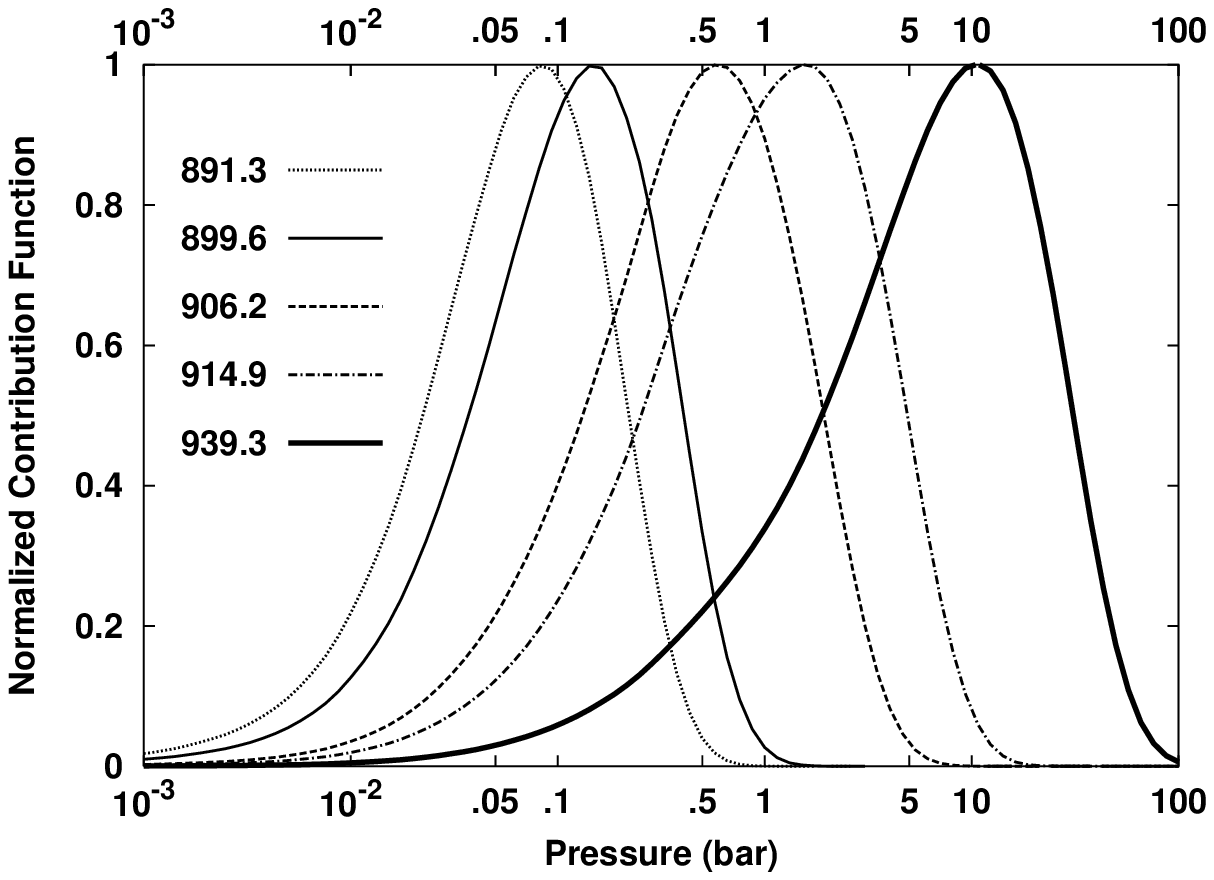}
\caption{(a)(Top): Distribution of contribution functions in the 727-nm
 data set. (b)(Bottom): Distribution of contribution functions in the
 890-nm data set. The listed wavelengths are all in units of nm.}  
\end{center}
\end{figure}

To analyze the center-limb profiles, we employ a forward modeling
approach. We construct a cloud structure model and compute synthetic
center-limb profiles at the selected wavelengths using a radiative
transfer code. The model profile in a strong absorption band should be
affected only by the uppermost part of the model, while that in a weaker
band or continuum is affected by a broad altitude range of the model
structure. Thus, the model that can simultaneously reproduce a set of
different observed profiles within the calibration uncertainty is
considered to represent the vertical distribution of aerosols in
Saturn's atmosphere.

\section{Modeling}

\subsection{Modeling Overview and Uniqueness of our Trial}

\begin{figure}
\begin{center}
\scalebox{.75}{\includegraphics{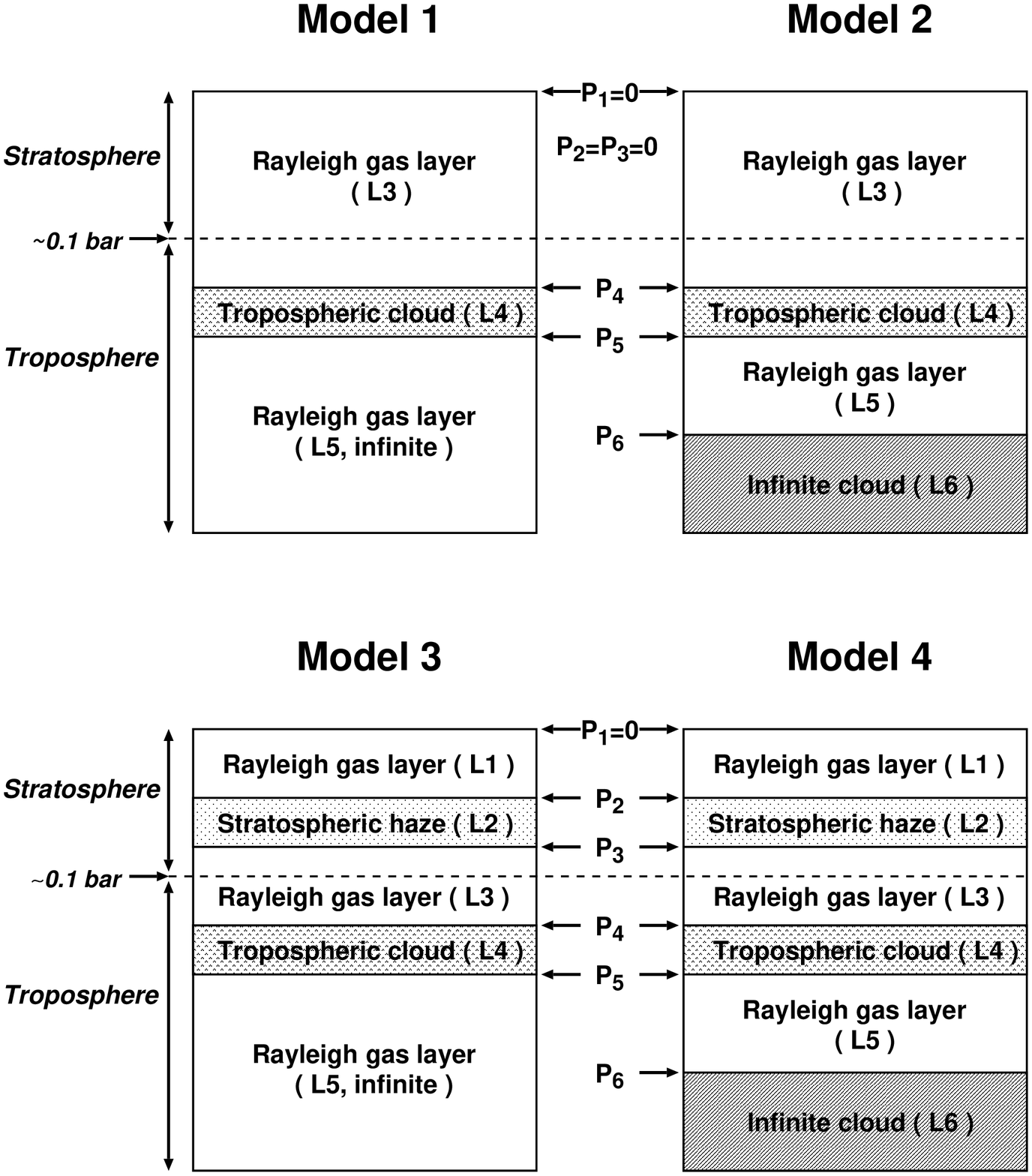}}
\caption{Four different cloud models used in our modeling process.}
\end{center}
\end{figure}

When we prepared our cloud structure models, we paid attention to the
lessons learned from the current debate about the existence of a thick
tropospheric cloud in the Jovian atmosphere. Before the entry of the
Galileo Probe into the Jovian atmosphere, researchers commonly assumed
the existence of a cloud of ammonium hydro-sulfide or water in the lower
Jovian troposphere (at 2--5 bar level) in their cloud structure modeling,
relying on the thermo-chemical prediction by Weideschilling and Lewis
(1973) and infrared data analyses (Marten {\it et al.} 1981, B\'ezard
{\it et al.} 1983, Carlson {\it et al.} 1993,1994). Nonetheless, the
Galileo Probe discovered a strong depletion in water in comparison with
the solar abundance ratio and detected no significant cloud opacity
below 2-bar  level during its decent in a region known as a 'hot
spot'. Although these hot spots are known to be relatively dry and
cloud-free, this surprising fact called into question the presence of a
thick cloud in the Jovian lower troposphere. Recently, Sromovsky and Fry
(2002) pointed out that it is possible to reproduce the optical
center-limb profiles of the Jovian hot spots without any cloud below
1.2-bar level. Since modeling approaches with a thick cloud placed at
the bottom were commonly adopted in previous Saturnian center-limb
analyses, we now think it is necessary to try new type of Saturnian
cloud models without the thick bottom cloud. 

We thus set up four different types of cloud structure models, as shown
in Fig. 5. These represent the existence of a stratospheric haze layer
(models 3 and 4) or the no-haze case (models 1 and 2), as well as the
existence of a lower infinite cloud (models 2 and 4) or the
no-infinite-cloud case (models 1 and 3). The most significant point to
consider is whether or not an infinite cloud at the bottom is needed in
the deep atmosphere to reproduce the observed profiles. There are at
most six layers within these four cases; the top Rayleigh gas layer
($\equiv$ Layer 1, L1), a stratospheric haze layer ($\equiv$ Layer 2,
L2), a middle Rayleigh gas layer ($\equiv$ Layer 3, L3), an upper
tropospheric cloud layer ($\equiv$ Layer 4, L4), a lower Rayleigh gas
layer ($\equiv$ Layer 5, L5) and a lower infinite cloud layer ($\equiv$
Layer 6, L6). All the Rayleigh gas layers are aerosol-free. The
stratospheric haze layer and upper cloud layer can be either of aerosols
only or a mixture of Rayleigh gas and aerosols, depending on the initial
assumptions and modeling results. The bottom infinite cloud does not
contain any Rayleigh gas component. For simplicity, we sometimes refer
to the stratospheric haze as 'haze', and the upper and lower
tropospheric clouds as 'UCLD' and 'LCLD', respectively.        

In these models, $P_1 (\equiv 0)$ is the pressure level of the
 atmosphere top. $P_2$ and $P_3$ signify the levels of the stratospheric
 haze top and bottom, respectively. $P_4$ and $P_5$ denote the levels of
 the top and bottom of the upper tropospheric cloud, and $P_6$ is
 defined as the top of the lower infinite cloud. The pressure difference
 between the top and bottom of a layer is denoted as $DLP$. For example,
 $DLP1$ is the pressure difference between the top and bottom of L1, so
 $DLP1 \equiv P_2 - P_1$. Similarly, $DLP2 \equiv P_3 - P_2$, $DLP3
 \equiv P_4 - P_3$, $DLP4 \equiv P_5 - P_4$ and $DLP5 \equiv P_6 -
 P_5$. In the actual modeling computation, these $DLP$ values are
 treated as free variables rather than the $P$ values.  

In models 1 and 2, we do not assume the existence of the
stratospheric haze, hence $P_1 = P_2 = P_3 = 0$. In models 1 and 3,
the lower Rayleigh gas layer (L5) is assumed to be infinite ($DLP5 =
9999.$). The LCLD in models 2 and 4 is given infinite aerosol optical
thickness ($\tau_{lcld} = 999$). Furthermore, we later sub-divide these
four models according to the types of scattering phase functions
employed in different aerosol layers. Among those models, we try to find
the one that can best reproduce the observed center-limb profiles within
our calibration uncertainty. The theory to calculate the optical
thickness and effective single-scattering albedo of each layer is
summarized in Appendix B.    
 
We point out three advantages of our data acquisition and modeling
approach. Since the AOTF camera provides 2D images at higher spectral
resolution around methane bands than that of traditional narrow-band
filters, we are able to investigate the entire southern hemisphere of
Saturn with higher vertical sampling resolution than before. Moreover,
the wide spectral range of AImS enables us to study the cloud structure
at different absorption bands. This allows us to examine the wavelength
dependence of the optical aerosol properties, yielding important
information concerning the size distribution and composition of Saturn's
aerosols. In addition, our computational algorithm constrains free model
parameters more rigorously than before. Previously, center-limb profiles
of different wavelengths were fit one by one, adjusting only a few
variables of strong influence on the chosen profile ({\it e.g.} varying
only aerosol albedos in continuum fit). Obviously, this technique
ignores the mutual dependencies among different variables. In contrast,
we do not have this problem because our code simultaneously fits a
set of selected profiles and returns a complete set of several fit
parameters.

\subsection{Selection of Gas and Aerosol Scattering Phase Functions}

There is some flexibility in the choice of scattering phase functions
for the aerosol layers. Typically, the two-term Henyey-Greenstein
function (hereafter, TTHGF) derived from the Pioneer observations is
used to simulate the scattering of the thick tropospheric cloud of
Saturn (Tomasko and Doose 1984, Karkoschka and Tomasko 1992, Ortiz {et
al.} 1996). This function has the following form:   

\begin{equation}
P(\theta)= f \cdot P(g_1,\theta) + (1-f) \cdot P(g_2,\theta)
\end{equation}

and

\begin{equation}
P(g,\theta)= \frac{1-g^2}{(1+g^2 - 2g\cos\theta)^{\frac{3}{2}}}
\end{equation}

where $\theta$ denotes scattering angle, $f$ (0$\leq f \leq$1) is the
fraction of forward-scattered light, $g_1$ (0$\leq g_1 \leq$1) is the
asymmetry factor of forward-scattering and $g_2$ (-1$\leq g_2 \leq$0) is
the asymmetry factor of back-scattering. A larger value of $g_1$ gives a
sharper forward-scattering peak, while a larger absolute value of $g_2$
gives a sharper back-scattering peak. The TTHGF obtained by Tomasko and
Doose (1984) in red light in the southern Saturnian equatorial
region (from $-7^{\circ}$ to $-11^{\circ}$) ([$f,g_1,g_2$]=[0.763,
0.620, -0.294]) is used as the scattering phase function of the lower
tropospheric cloud throughout our modeling. For the stratospheric haze
and UCLD, we use both the above TTHGF of Tomasko and Doose (1984) and Mie
scattering phase functions of different aerosol size distributions. The
modified gamma function (Hansen and Travis 1974), which is defined by
two parameters, effective radius $a$ and effective variance $b$, was
employed to simulate the particle size distribution. The average Mie
phase function for a given size distribution was obtained by weighting
and adding up Mie phase functions of different aerosol sizes. For the
purpose of Mie computation for a certain particle size, we adopted the
code of Bohren and Hoffman (1983). The adopted parameters in the Mie
scattering computation are listed in Table 1.    

\begin{table} 
\begin{center}
\begin{tabular}{ccccc} \hline
Applied layer & $a$ & $b$  & $n_r$ \\ \hline
Stratospheric haze       & 0.15 & 0.1 & 1.43 \\ 
Upper tropospheric cloud & 1.50 & 0.1 & 1.43 \\ \hline
\end{tabular}
\caption{List of Mie scattering parameters used in our modeling. $a$ is
 the effective aerosol radius in $\mu$m, and $b$ is the effective
 variance of Hansen's size distribution function (Hansen and Travis
 1974). $n_r$ represents the real refractive index of aerosols.}   
\end{center}
\end{table}

The values listed in Table 1 were employed because they were used in
previous publications (Karkoschka and Tomasko 1993, Ortiz {\it et al.}
1996) and facilitate the comparison between our results and previous
work. The photochemically expected condensates in the Saturnian
stratosphere include diacetylene ($C_4H_2$, $n_r=1.42$) and ethane
($C_2H_6$, $n_r=1.44$) (Karkoschka and Tomasko 1993). Therefore we adopt
the average of those indices as the refractive index for our haze
layer. The refractive index of the UCLD is assumed to be that of ammonia
ice. We assume the imaginary part of aerosol refractive index ($n_i$) is
zero. Instead, absorption by aerosols is taken into account by adjusting
the single-scattering albedo of aerosols in the modeling process. With
these particle parameters, we generated the Mie phase functions at 730
and 900 nm, and applied them to the 727-nm data set and the 890-nm data
set, respectively. 

At this point, we sub-divided the 4 main models into 12 models,
depending on the phase functions adopted in the haze and UCLD layers
(Table 2). We refer to the adopted TTHGF as 'H-G (red)', the Mie
function for the haze as 'Mie 0.15' and the Mie function for the UCLD as
'Mie 1.5'.         

\begin{table}
\begin{center}
\caption{Cloud models with different phase functions ($P(\theta)$)} 
\vspace{3mm}

\begin{tabular}{|c|c|c|c|c|} \hline
Parent Model & Model \# & Haze $P(\theta)$ & UCLD $P(\theta)$ & LCLD $P(\theta)$ \\ \hline
1 & 1-1 & N.A. & H-G (red) & H-G (red) \\ \cline{2-5}
  & 1-2 & N.A. & Mie 1.5   & H-G (red) \\ \hline\hline
2 & 2-1 & N.A. & H-G (red) & H-G (red) \\ \cline{2-5}
  & 2-2 & N.A. & Mie 1.5   & H-G (red) \\ \hline\hline
  & 3-1 & H-G (red) & H-G (red) & H-G (red) \\ \cline{2-5}
3 & 3-2 & H-G (red) & Mie 1.5   & H-G (red) \\ \cline{2-5}
  & 3-3 & Mie 0.15  & H-G (red) & H-G (red) \\ \cline{2-5}
  & 3-4 & Mie 0.15  & Mie 1.5   & H-G (red) \\ \hline\hline
  & 4-1 & H-G (red) & H-G (red) & H-G (red) \\ \cline{2-5}
4 & 4-2 & H-G (red) & Mie 1.5   & H-G (red) \\ \cline{2-5}
  & 4-3 & Mie 0.15  & H-G (red) & H-G (red) \\ \cline{2-5}
  & 4-4 & Mie 0.15  & Mie 1.5   & H-G (red) \\ \hline
\end{tabular}
\end{center}
\end{table}

The scattering phase function in the Rayleigh gas layers is that of
Rayleigh scattering ($P_R(\theta)$):

\begin{equation} 
P_R(\theta) = \frac{3}{4}(1 + cos^2 \theta)
\end{equation}

where $\theta$ again denotes scattering angle. In the cases where
aerosols and Rayleigh gas co-exist in a layer, we assume that they are
uniformly mixed and adopt the aerosol's phase function since the opacity
by the Rayleigh scattering ($\tau_R$) was always negligibly small
compared with the aerosol opacity ($\tau_A$) due to the
relatively long wavelength of our observations in any layer ($\tau_R /
\tau_A < 10^{-3}$). All the phase functions adopted in our modeling are
illustrated in Fig. 6.  

\begin{figure}
\begin{center}
\includegraphics{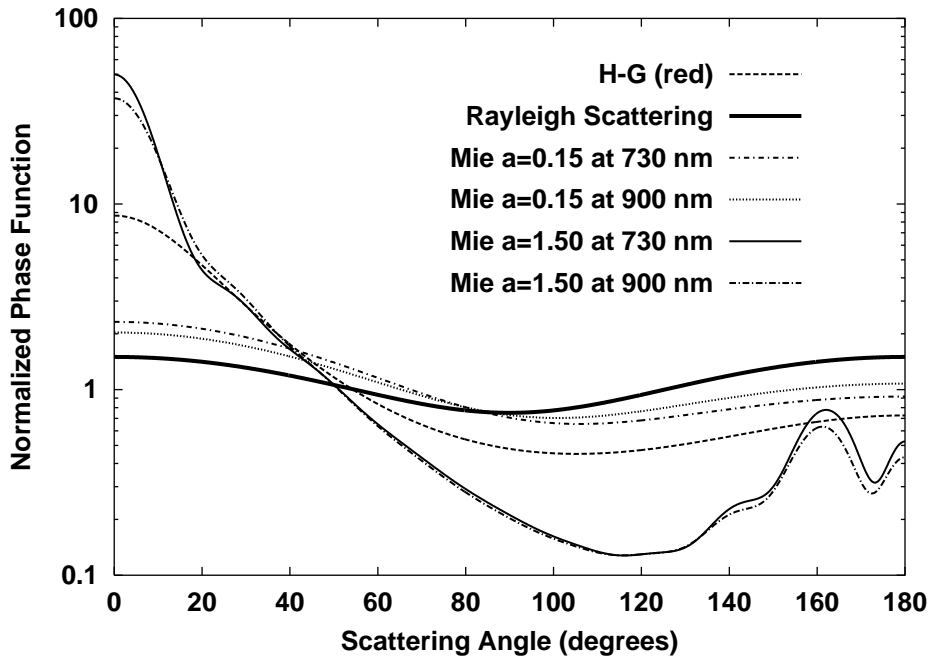}
\caption{Different phase functions used in our modeling.}  
\end{center}
\end{figure}

\subsection{Atmospheric Composition and Methane Absorption} 

With respect to the atmospheric composition, a number of different
results have been published (Hanel {\it et al.} 1981,
Conrath {\it et al.} 1984, Courtin {\it et al.} 1984, Atreya {\it et
al.} 1999, Conrath and Gautier 2000, Atreya {\it et al.} 2003). We chose
the values of Hanel {\it et al.} (1981) for the Saturnian hydrogen and
helium abundance and assumed that the methane mole fraction is 0.002
(Table 3). These are identical to those used by Stam {\it et al.}
(2001), and nearly equal to those used in other cloud modeling efforts
(Tomasko and Doose 1984, Karkoschka and Tomasko 1992, 1993, Ortiz {\it
et al.} 1996, Acarreta and S\'anchez-Lavega 1999). Therefore, this
choice also facilitates comparison between our results and previous
studies.            

\begin{table}
\begin{center}
\caption{Atmospheric composition of Saturn assumed in our modeling.}
\vspace{3mm}

\begin{tabular}{cc} \hline
Gas Species & Mole fraction \\ \hline
$H_2$ & 0.940  \\
$He$ & 0.058  \\
$CH_4$ & 0.002 \\ \hline
\end{tabular}
\end{center}
\end{table}

There are several publications that address the methane
absorption coefficients. Through the analysis of the spectra of giant
planets, Karkoschka (1994) reported that the optical methane band
shapes are narrower and band peaks are deeper at colder temperatures in
giant planets' atmospheres than at room temperatures. The newest
laboratory measurements of methane absorption coefficients at 77 K 
(Singh and O'Brien 1995, O'Brien and Cao 2002) showed very
good agreement with Karkoschka's result. Unfortunately, the data sets of
those new laboratory measurements do not fully cover the wavelength
ranges in either our 727- or 890-nm data sets. Therefore, we employ the
methane coefficient values derived by Karkoschka (1994) in this
paper. In our computations, these values were convolved with the
transmission function of the AOTF filter to obtain the effective value
at each selected passband (Table 4).          

\begin{table}
\begin{center}
\caption{Effective methane absorption coefficients ($k_{\mbox{\tiny $CH_4$}}$ in [km--am$]^{-1}$) for our wavelengths of interest ($\lambda$ in [nm]).}  
\vspace{3mm}

\begin{tabular}{cc} \hline
$\lambda$ & $k_{\mbox{\tiny $CH_4$}}$ \\ \hline
 726.5 & 3.863 \\
 732.1 & 1.653 \\
 735.8 & 0.690 \\
 747.4 & 0.015 \\ \hline
 891.3 & 24.471 \\     
 899.6 & 12.699 \\
 906.2 & 2.277 \\
 914.9 & 0.860 \\
 939.3 & 0.021 \\ \hline
\end{tabular}
\end{center}
\end{table}

\subsection{Free and fixed parameters}

The following 11 parameters are left undetermined:   

\begin{itemize}
\item Aerosol opacities in L2 ($\equiv \tau_{haze}$) and L4
      ($\equiv \tau_{ucld}$)    
\item Aerosol single-scattering albedos in L2 ($\equiv \varpi_{0,haze}$), L4 ($\equiv \varpi_{0,ucld}$) and L6 ($\equiv \varpi_{0,lcld}$)  
\item Rayleigh scattering albedo at continua
\item Physical thickness of L1 -- L5 ($DLP1$ -- $DLP5$).  
\end{itemize}

These parameters must be fixed by further assumptions or determined from
the center-limb profile fitting. We set the continuum Rayleigh
scattering albedo to 1.0, considering no gas absorption other than that
of methane. In addition, we fixed the physical thickness of the haze
layer. According to Karkoschka and Tomasko (1993), the optical thickness
of the stratospheric haze is always very small, near $0.2$ at 340 nm,
when the pressure difference between the top and bottom of the
stratospheric haze ($DLP2$) is roughly 100 mb. In the spectral region of
our data sets (726 -- 940 nm), this opacity would be even smaller when
we consider the likely particle size ($\sim 0.15$ $\mu$m) of the haze
(Karkoschka and Tomasko 1993). From this argument, we assume that
variation in $DLP2$ has only a minor effect on the computation result
and therefore adopt the value 10 mb for $DLP2$. All the other pressure
variables are left as free variables; the danger of assuming a certain
pressure level for any cloud deck is well explained by Sromovsky and Fry
(2002) and others for the case of the Jovian atmosphere.       

In the end, we are left with at most 9 free variables ($\tau_{haze}$,
  $\varpi_{0,haze}$, $\tau_{ucld}$, $\varpi_{0,ucld}$,
  $\varpi_{0,lcld}$, $DLP1$, $DLP3$, $DLP4$, $DLP5$) to be adjusted to
  reproduce the observed center-limb profiles.

\subsection{Optimization procedure and initial conditions}

\subsubsection{Optimization procedure}

To examine the wavelength dependence of aerosol opacity and albedo, our
 fitting operations for the 727- and 890-nm data sets were done
 individually. First, we fit the 890-nm data set to determine the entire
 cloud structure because the 890-nm data set more strictly constrains
 the model with the wider altitude coverage and more wavelengths. Then,
 from the best-fitting parameters of the 890-nm data set, we adopt the
 pressure levels of aerosol layers and fix these parameters in the
 subsequent 727-nm data set simulations. We then obtain aerosol
 opacities and albedos near 727 nm. Thus, in addition to the cloud
 structure, we can determine the wavelength dependence of optical
 properties of aerosols.       
 
The intensity computation was done with a radiative transfer code based
on the adding-doubling method (Hansen 1969). The accuracy of fitting for
each data set is judged by the following reduced $\chi^2$.   

\begin{equation}
 \chi^2 = \frac{1}{N_{free}} \sum_i \displaystyle{\frac{((I/F)_{i,obs}-(I/F)_{i,com})^2} {\sigma_{i,obs}^2}}
\end{equation}

where

\begin{equation}
  N_{free} = N_{points}-N_{variable},
\end{equation}

\begin{description}
 \item[$(I/F)_{i,obs}$ :] Observed $I/F$ value at {\it i} th data point,
 \item[$(I/F)_{i,com}$ :] Computed $I/F$ value at {\it i} th data point,
 \item[$\sigma_{i,obs}$ :] Observational error of $I/F$ at {\it i} th data point,
 \item[$N_{free}$ :] Degree of freedom of the fit,  
 \item[$N_{points}$ :] Number of data points of a profile,
 \item[$N_{variable}$ :] Number of variables used in fitting a profile.
\end{description}

Numerically, a solution is considered acceptable if the $\chi^2$ is
smaller than 1.0.  

We search multi-dimensional parameter space (up to 9 dimensions) for the
 optimum combination of parameter values that can simultaneously fit a
 certain number of profiles in a given data set (five in the 890-nm data
 set and four in the 727-nm data set) within our calibration error. This
 is equivalent to seeking the minimum point of $\chi^2$. To do this, our
 code follows the gradient-expansion method in Bevington and Robinson
 (1992). Given an initial set of parameters, this method computes the
 gradient of $\chi^2$ at that point in the multi-variable space, and
 takes the next point of smaller $\chi^2$ along the gradient
 vector. This process is iterated until a 'hollow' of $\chi^2$ is
 found. However, the well-known difficulty is that the discovered
 hollow point may be just one of the local-minimum points of
 $\chi^2$, regardless of the adopted optimization method. The
 only way to avoid this concern is to cover the entire parameter space
 with a fine grid and adopt grid-search method, though the huge
 computational burden makes this approach impractical. Therefore, a good
 initial setting is essential to reach the correct minimum point in our
 modeling. Nevertheless, since there is not enough information available to
 make reasonable presumptions about desirable initial parameters, our
 initial settings must cover as wide a range as possible in the
 parameter space. Due to this requirement, we integrated the
 grid-search algorithm into our code so that our initial conditions can
 systematically explore the vast parameter space within their physically
 reasonable range. Namely, we set up a grid in the multi-parameter space
 and let all the grid-points represent our initial conditions for
 optimization. 

In practice, for a chosen type of model, we specify a large number of
 combinations of reasonable initial values. Starting from each of those
 initial conditions, our computer program finds a local minimum point of
 $\chi^2$ and records it as a solution for the given initial
 condition. Among those recorded solutions, we choose the one that gives
 the smallest $\chi^2$ for the chosen cloud model. The same procedure is
 used for all the different cloud models to obtain a model-specific
 solution for each cloud model. After examining and comparing those
 model-dependent solutions, we finally choose the true best-fit result
 for the given data set of Saturn.

\subsubsection{Initial conditions and their limitations}

\begin{table}
\begin{center}
\caption{The minimum, maximum and initial values of free variables. All
 the $DLP$ values are in units of {\it bar}.} 
\vspace{3mm}

\begin{tabular}{|c|c|c|c|} \hline
 Free Variable (if it exists) & Min. & Max. & Initial values \\ \hline
$\tau_{haze}$     & 0.   & 0.5  & 0.1 \\ \hline
$\tau_{ucld}$     & 0.   & 50.  & 1., 5., 10., 30. \\ \hline
$\varpi_{0,haze}$ & 0.95 & 1.0  & 0.975, 0.995 \\ \hline
$\varpi_{0,ucld}$ & 0.95 & 1.0  & 0.975, 0.995 \\ \hline
$\varpi_{0,lcld}$ & 0.95 & 1.0  & 0.975, 0.995 \\ \hline
$DLP1$            & 0.   & 0.1  & 0.05 \\ \hline
$DLP3$            & 0.   & 10.  & 0.1, 0.5, 1., 3. \\ \hline
$DLP4$            & 0.   & 20.  & 0.1, 0.5 \\ \hline
$DLP5$            & 0.   & 9999.& 0.1, 0.5, 1. 3. \\ \hline
\end{tabular}
\end{center}
\end{table}

The minimum, maximum and initial values of the free variables in our
modeling are summarized in Table 5. The maximum value of $\tau_{haze}$
is set to 0.5 because the haze opacity should be small based on the
argument of Karkoschka and Tomasko (1993). $\tau_{ucld}$ is one of the
most important parameters in our modeling. Therefore, we allow it to
have a large range of possible values. The maximum of $\tau_{ucld}$ is
practically infinite ($\tau_{ucld}=50.$), and four different initial
values are given spanning 1 -- 30. In light of previous modeling results
(Tomasko and Doose 1984, Karkoschka and Tomasko 1992,1993, Ortiz {\it et
al.} 1996, Acarreta and S\'anchez-Lavega 1999), all aerosol albedos
($\varpi_{0,haze}$, $\varpi_{0,ucld}$ and $\varpi_{0,lcld}$) are limited
between 0.95 and 1.0. They are assigned one of two possible initial
values, 0.975 and 0.995, to represent either an absorbing or a
reflective case.      

$DLP1$ controls the altitude of the stratospheric haze. Hence, its upper
limit is 0.1 bar, which corresponds to the entire stratosphere. The gap
between the haze and UCLD is given by $DLP3$. Here, we chose the maximum
value of 10 because $DLP3$ value cannot exceed this and still reproduce
the observed intensity at strong absorption bands when there is only a
thin haze layer above the UCLD. Since this is an important parameter for
the 727-nm data fit, we use four initial values. Another important
parameter, especially in the model group 1, is the thickness of the
UCLD, $DLP4$. The maximum of this variable is allowed to be 20, which
includes the extreme case where the UCLD is an aqueous $H_2O-NH_3$ cloud
extending up to the stratosphere. Only two small initial values are used
for $DLP4$ because we initially suppose that the UCLD is not physically
very thick. The vertical 'distance' between the UCLD and LCLD is
determined by $DLP5$. The maximum is set to 9999., which means there is
no LCLD. The four initial values of $DLP5$ take into consideration all
the cases where the LCLD has a major and minor contribution to the
outgoing radiation.    

In the model group 1, the free parameters are $\tau_{ucld}$,
$\varpi_{0,ucld}$, $DLP3$ and $DLP4$. Thus, 64 (= $4 \times 2 \times 4
\times 2$) types of initial conditions are used. In the model group 2,
the free variables are $\tau_{ucld}$, $\varpi_{0,ucld}$,
$\varpi_{0,lcld}$, $DLP3$, $DLP4$ and $DLP5$. They have 512 (= $4 \times
2\times 2 \times 4 \times 2 \times 4$) types of initial value sets. In
the model group 3, $\tau_{haze}$, $\tau_{ucld}$, $\varpi_{0,haze}$,
$\varpi_{0,ucld}$, $DLP1$, $DLP3$ and $DLP4$ are adjusted. This
corresponds to 128 (= $1 \times 4 \times 2 \times 2 \times 1 \times 4
\times 2$) types of initial conditions. In the model group 4, all the
nine parameters in Table 5 are adjusted to fit the profiles, and a total
of 1024 (= $1 \times 4 \times 2 \times 2 \times 2 \times 1 \times 4
\times 2 \times 4$) types of initial conditions are tested. In the end,
we explored a total of 5760 different initial conditions for the 890-nm
data fitting and obtain the same number of corresponding solutions. The
subsequent 727-nm data fitting used a total of 240 initial
conditions. The entire computation process took about 70 hours on a 2.8
GHz Pentium workstation. By examining those numerous solutions, we
determined the best-fit cloud model for Saturn's equatorial region over
726 -- 940 nm.       

\subsection{Radiative Transfer code check}

Our adding-doubling code was originally written for the analysis of
Pioneer 10 and 11 data. It has a total of 22 quadrature points for the
integration over the scattering angles and can hold up to 60 Fourier
azimuthal expansion terms. We checked the accuracy of this code using
several types of scattering phase functions.     

\subsubsection{Simple phase functions}

First, we tested the computation accuracy using simple phase
functions. The simplest isotropic scattering case was tested in
comparison with the analytic solutions of Chandrasekhar (1960). A
Rayleigh scattering case was 
also examined following the method of Stammes {\it et al.} (1989). With
the phase functions of Jupiter's cloud and haze, we cross-matched our
computational intensities with those of Dr. T. Satoh (private
communication). Throughout these procedures, the agreement was excellent
($\leq 0.5\%$) and therefore we are confident that our simulation is
sufficiently accurate when the above phase functions are employed. We also
compared our results with those obtained using the DISORT code that is
based on the Discrete Ordinate Method (Stamnes {\it et al.} 1988, Thomas
and Stamnes 1999). This method is conceptually very different from the
Adding-Doubling method to solve radiative transfer problems, as
concisely described by Hansen and Travis (1974). In the DISORT
computation, 3 streams were taken for the isotropic and Rayleigh
scattering cases, and 32 or higher was used for other cases. Again,
the agreement was nearly perfect (deviation $\leq 0.1\%$) with 
the use of the above phase functions and the Saturnian TTHGFs.         

\subsubsection{Mie phase function}

Our concern regarding the Mie scattering case was that the sharp forward
 scattering lobes seen in Mie phase functions would not be accurately
 included in our computation. Benassi {\it et al.} (1984) was referred
 to for the accurate computational results with highly asymmetric Mie
 phase functions.     

The accuracy of our computations deteriorated when the effective aerosol
radius reached about 10 $\mu$m and the corresponding forward scattering
lobe became too sharp (the normalized functional value is $10^3$ --
$10^4$ at $\theta = 0^{\circ}$). This is probably because the quadrature 
points of our code are not distributed finely enough to accurately
integrate a very sharp forward scattering peak. On these occasions we
followed the truncation method of Potter (1970), in which the
forward-scattered light in the diffraction peak is regarded as
unscattered. This operation results in a significant reduction in
scattering cross section of aerosols, and therefore we need to
appropriately re-scale aerosol opacities and albedos. This truncation
dramatically improved our computational accuracy in the large particle
cases. The deviation from the published values in Benassi {\it et al.}
(1984) was less than 1\% in most cases, though there were a few cases
where the deviations were as large as $\sim 3\%$.    

Fortunately, with the particle sizes employed in our Saturnian modeling
 computations ($a$ = 0.15 and 1.5), no truncation was necessary
 because the Mie phase functions were smooth enough in those
 cases. Compared with a 64-stream DISORT result, the computational error
 was smaller than 0.06\% for $a$=0.15 and smaller than 0.3\%
 for $a$=1.5. These deviations are totally negligible in our modeling.

\section{Results and Discussion}

The fitting accuracies for all of our models are shown in Table 6.   

\begin{table}
\begin{center}
\caption{Equatorial profile fitting result}
\vspace{3mm}

\begin{tabular}{|c|c|c|c|} \hline
Model & Best 5 $\chi^2$ in 890-nm data set & Best 5 $\chi^2$ at
 939.3 nm & Best $\chi^2$ in 727-nm data set \\ \hline
1-1 & 0.177 -- 0.193 & 0.33 -- 0.43 & 0.107 \\ \hline
1-2 & 0.402 -- 0.441 & 0.86 -- 1.12 &    -  \\ \hline\hline 
2-1 &       0.149    &       0.30   & 0.437 \\ \hline
2-2 & 0.297 -- 0.298 & 0.80 -- 0.81 &    -  \\ \hline\hline 
3-1 & 0.192 -- 0.240 & 0.34 -- 0.55 & 0.082 \\ \hline
3-2 & 0.261 -- 0.488 & 0.45 -- 0.93 &    -  \\ \hline
3-3 & 0.186 -- 0.299 & 0.32 -- 0.95 & 0.079 \\ \hline
3-4 & 0.222 -- 0.313 & 0.39 -- 0.65 &    -  \\ \hline\hline 
4-1 & 0.151 -- 0.153 & 0.30 -- 0.31 & 0.251 \\ \hline
4-2 & 0.179 -- 0.185 & 0.40 -- 0.41 &    -  \\ \hline
4-3 & 0.139 -- 0.144 & 0.23 -- 0.25 & 0.098 \\ \hline
4-4 & 0.160 -- 0.162 & 0.29 -- 0.31 & 0.095 \\ \hline
\end{tabular}
\end{center}
\end{table}

\subsection{890-nm data set fitting}

In our fitting of the 890-nm data set, we obtained a number of different
solutions derived from various initial conditions based on our 
cloud models. The $\chi^2$ values for the entire data set are well
smaller than 1 and very close to one another for each assumed model,
making it difficult to choose the best among them if we use the $\chi^2$
alone. Therefore, we first selected the five best solutions for each
model. The main deviation source of this 890-nm data set is the
flattened profile shape from the central meridian towards the dusk limb
(of negative relative longitude) of the 939.3-nm profile, as shown in
Fig. 7. This discrepancy becomes more prominent as wavelength increases
from 891.3 nm (Fig. 8). As a result, the fitting residual of this 939.3-nm
profile dominates the overall fitting accuracy. For this reason, we use
the $\chi^2$ value at 939.3 nm as an additional criterion for assessing
the fitting accuracy for this data set. In Table 6, the degradation of
939.3-nm fit is obvious when the UCLD adopts the 'Mie 1.5' phase
function. In that case, the reproduced limb-darkening becomes steep and
the computed intensity tends to be low at the both limbs, presumably
because of the strong forward scattering of the UCLD. This trend is the
most evident when there is no haze layer above the UCLD ({\it e.g.}
models 1-2 and 2-2). The comparison between the results of model 1-1 and
1-2 clearly illustrates this effect (Fig. 7). Therefore, the average
tropospheric particle size smaller than 1.5 $\mu$m is preferred in our
models.       

\begin{figure}
\begin{center}
\includegraphics{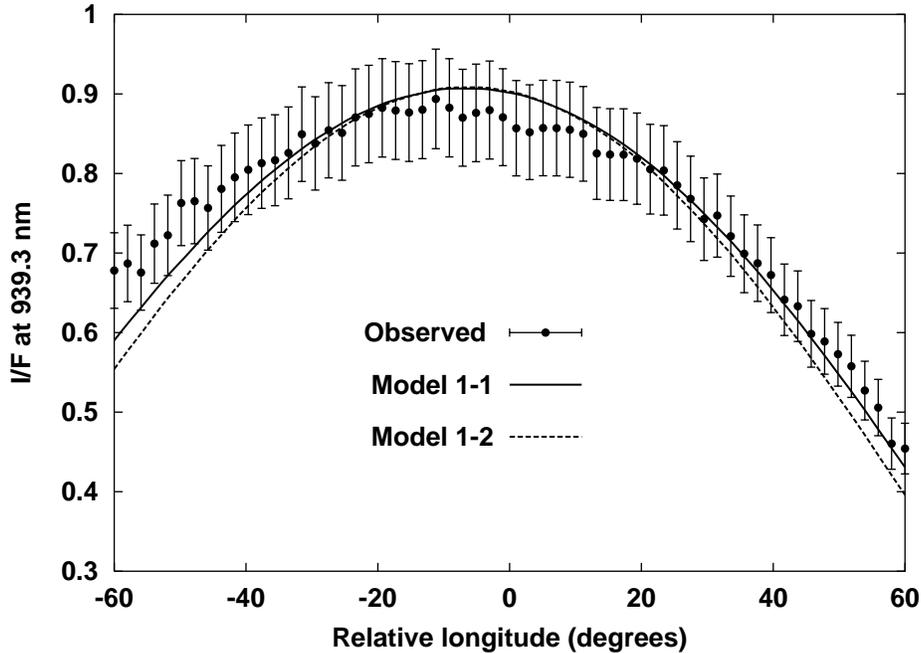}
\caption{Plot of the 939.3-nm profile and comparison of best-fit results
 at 939.3 nm between model 1-1 and 1-2. The error bars show $\pm$7\% of
 calibration error.}   
\end{center}
\end{figure}

To proceed to the 727-nm data fitting, we selected only the models that
satisfy the following two conditions : i) $\chi^2$ at 939.3 nm must be 
smaller than 0.4, ii) $\chi^2$ for overall 890-nm data set must be
smaller than 1.2 times the best $\chi^2$ in the model group to which the 
considered model belongs. The first condition is adopted because the
systematic deviation between the computed and observed 939.3-nm profiles
becomes visually clear when this value exceeds 0.4. The second
condition is set rather empirically and arbitrarily as a constraint on
the fitting at the other wavelengths. This 20\% range of $\chi^2$
corresponds to the difference in the best $\chi^2$ between models
3-3 and 3-4 or between models 4-2 and 4-3. The reason we apply this
second condition only within each model group is that we examine the
dependence of solutions on assumed cloud models. As a consequence of the
above two criteria, the solutions from models 1-1, 2-1, 3-1, 3-3, 4-1,
4-3 and 4-4 remain qualified as best-fit for Saturn's equatorial
atmospheric structure near 890 nm. Next, we fit the 727-nm data set
using these qualified cloud models.       

\subsection{727-nm data set fitting}

We fit the 727-nm data set using the cloud layer altitudes of the
 qualified models obtained through the 890-nm data fitting. The best
 results are shown again in Table 6. In this data set, the dawn-dusk
 asymmetry is weaker than in the 890-nm data set, hence the smaller
 residual $\chi^2$ for the best-fit results.  

It is clear from an examination of Table 6 that models 2-1 and 4-1
should be discarded on account of their significantly poorer fitting
accuracy than the others. Consequently, seven solutions from five
different models (one solution from each of model 1-1, 4-3 and 4-4, and
two from each of 3-1 and 3-3) remain. These solutions are described in
Table 7, and displayed graphically in Fig. 9. The two solutions from
each of model 3-1 or 3-3 are derived from different initial
conditions. The uncertainties in Table 7 correspond to a 20\% of
increase in $\chi^2$. We allowed this $\chi^2$ range to be consistent
with the argument of criterion-setting in the preceding section. These
parameter uncertainties quantify how sensitively the fitting accuracy
depends on different variables.    

\begin{table}
\caption{Best-fit modeling results for the equatorial region of
 Saturn. (a)(Top): Solutions without LCLD. (b)(Bottom): Solutions with
 LCLD.} 
\vspace{5mm}

\begin{tabular}{|c||c|c|c|c|c|c|} \hline
Variables         &  1-1    & 3-1 (1) & 3-1 (2) & 3-3 (1) & 3-3 (2) \\ \hline
$\chi^2_{890}$    &  0.193  & 0.192  & 0.205 & 0.186  & 0.191 \\ \hline\hline
$\tau_{haze,890}$ &   -     & 0.00$+$0.18 & 0.21$\pm^{0.24}_{0.12}$ & 0.31$\pm^{0.19}_{0.05}$ & 0.27$\pm^{0.23}_{0.07}$ \\ \hline
$\tau_{ucld,890}$ & 16.97$\pm^{5.53}_{1.71}$ & 14.82$\pm^{4.97}_{1.00}$
 & 39.42$\pm^{10.56}_{3.43}$ & 47.66$\pm^{2.31}_{3.68}$ &
 49.81$\pm^{0.15}_{3.06}$ \\ \hline
$\varpi_{0,haze,890}$ &  -  & 1.000$-$0.050 & 1.000$-$0.018 & 1.000$-$0.010 & 1.000$-$0.012 \\ \hline
$\varpi_{0,ucld,890}$ & 0.998$\pm^{0.002}_{0.001}$ & 0.998$\pm$0.001 &
 0.997$\pm^{0.002}_{0.001}$ & 0.997$\pm^{0.002}_{0.001}$ &
 0.997$\pm^{0.002}_{0.001}$ \\ \hline  
$\varpi_{0,lcld,890}$ &  -  &  -  &  -  & - & - \\ \hline\hline
$\chi^2_{727}$ & 0.107 & 0.102 & 0.082 & 0.079 & 0.082 \\ \hline\hline 
$\tau_{haze,727}$ &  -  & 0.12$\pm^{0.09}_{0.11}$ & 0.01$\pm^{0.18}_{0.01}$ & 0.00$+$0.34 & 0.00$+$0.23 \\ \hline
$\tau_{ucld,727}$ & 8.87$\pm^{0.65}_{0.53}$ & 8.407$\pm^{0.42}_{0.45}$ &
 19.12$\pm^{1.54}_{3.94}$ & 19.55$\pm^{3.17}_{4.49}$ &
 27.03$\pm^{2.38}_{6.87}$ \\ \hline  
$\varpi_{0,haze,727}$ &  -  & 0.950$+$0.030 & 0.968$\pm^{0.032}_{0.018}$ &
 0.997$\pm^{0.003}_{0.047}$ & 0.989$\pm^{0.011}_{0.039}$ \\ \hline
$\varpi_{0,ucld,727}$ & 0.999$\pm$0.001 & 1.000$-$0.001 &
 0.997$\pm$0.001 & 0.996$\pm^{0.002}_{0.001}$ & 0.996$\pm$0.001 \\ \hline
$\varpi_{0,lcld,727}$ &  -  &  -  &  -  &  -  &  -  \\ \hline\hline
$DLP1$  &   -  & 0.004$\pm^{0.003}_{0.004}$ & 0.010$\pm^{0.002}_{0.009}$
 & 0.018$\pm^{0.002}_{0.008}$ & 0.021$\pm^{0.002}_{0.007}$ \\ \hline
$DLP3$  & 0.024$\pm^{0.004}_{0.007}$ & 0.010$\pm^{0.003}_{0.007}$ &
 0.014$\pm^{0.003}_{0.011}$ & 0.016$\pm^{0.002}_{0.014}$ &
 0.003$\pm^{0.002}_{0.003}$ \\ \hline 
$DLP4$  & 0.269$\pm^{0.031}_{0.072}$ & 0.240$\pm^{0.017}_{0.072}$ &
 0.515$\pm^{0.049}_{0.163}$ & 0.512$\pm^{0.043}_{0.174}$ & 0.719$\pm^{0.046}_{0.297}$ \\ \hline
$DLP5$  &  -  &  -  &  -  &  -  &  - \\ \hline
\end{tabular}

\vspace{1cm}

\begin{tabular}{|c||c|c|} \hline
Variables         &  4-3  &  4-4 \\ \hline
$\chi^2_{890}$    & 0.144 & 0.162\\ \hline\hline
$\tau_{haze,890}$ & 0.50$-$0.08 & 0.50$-$0.07 \\ \hline
$\tau_{ucld,890}$ & 5.17$\pm^{1.36}_{0.38}$    &  9.30$\pm^{2.08}_{0.80}$ \\ \hline
$\varpi_{0,haze,890}$ & 1.000$-$0.008  & 1.000$-$0.009 \\ \hline
$\varpi_{0,ucld,890}$ & 1.000$-$0.001 & 1.000$-$0.001 \\ \hline 
$\varpi_{0,lcld,890}$ & 0.993$\pm^{0.005}_{0.004}$ & 0.997$\pm$0.002 \\
 \hline\hline 
$\chi^2_{727}$ & 0.098 & 0.095 \\ \hline\hline 
$\tau_{haze,727}$ & 0.12$\pm^{0.10}_{0.08}$ & 0.14$\pm^{0.20}_{0.14}$ \\ \hline
$\tau_{ucld,727}$ &  4.22$\pm^{0.23}_{0.35}$ & 7.39$\pm^{0.39}_{0.50}$ \\ \hline 
$\varpi_{0,haze,727}$ & 0.950$+$0.036 & 1.000$-$0.025 \\ \hline
$\varpi_{0,ucld,727}$ & 1.000$-$0.001 & 1.000$-$0.001 \\ \hline
$\varpi_{0,lcld,727}$ & 0.994$\pm^{0.004}_{0.003}$ &
 0.995$\pm^{0.002}_{0.001}$ \\ \hline\hline 
$DLP1$  & 0.023$\pm^{0.002}_{0.008}$ & 0.022$\pm^{0.003}_{0.008}$\\ \hline
$DLP3$  & 0.024$\pm^{0.004}_{0.013}$ & 0.023$\pm^{0.004}_{0.012}$ \\ \hline
$DLP4$  & 0.026$\pm^{0.008}_{0.019}$ & 0.026$\pm^{0.008}_{0.017}$ \\ \hline
$DLP5$  & 0.567$\pm^{0.914}_{0.495}$ & 0.540$\pm^{0.840}_{0.422}$ \\ \hline
\end{tabular}
\end{table}

\begin{figure}
\begin{center}
\begin{tabular}{cc}
\begin{minipage}{80mm}
\begin{center}
\scalebox{.7}{\includegraphics{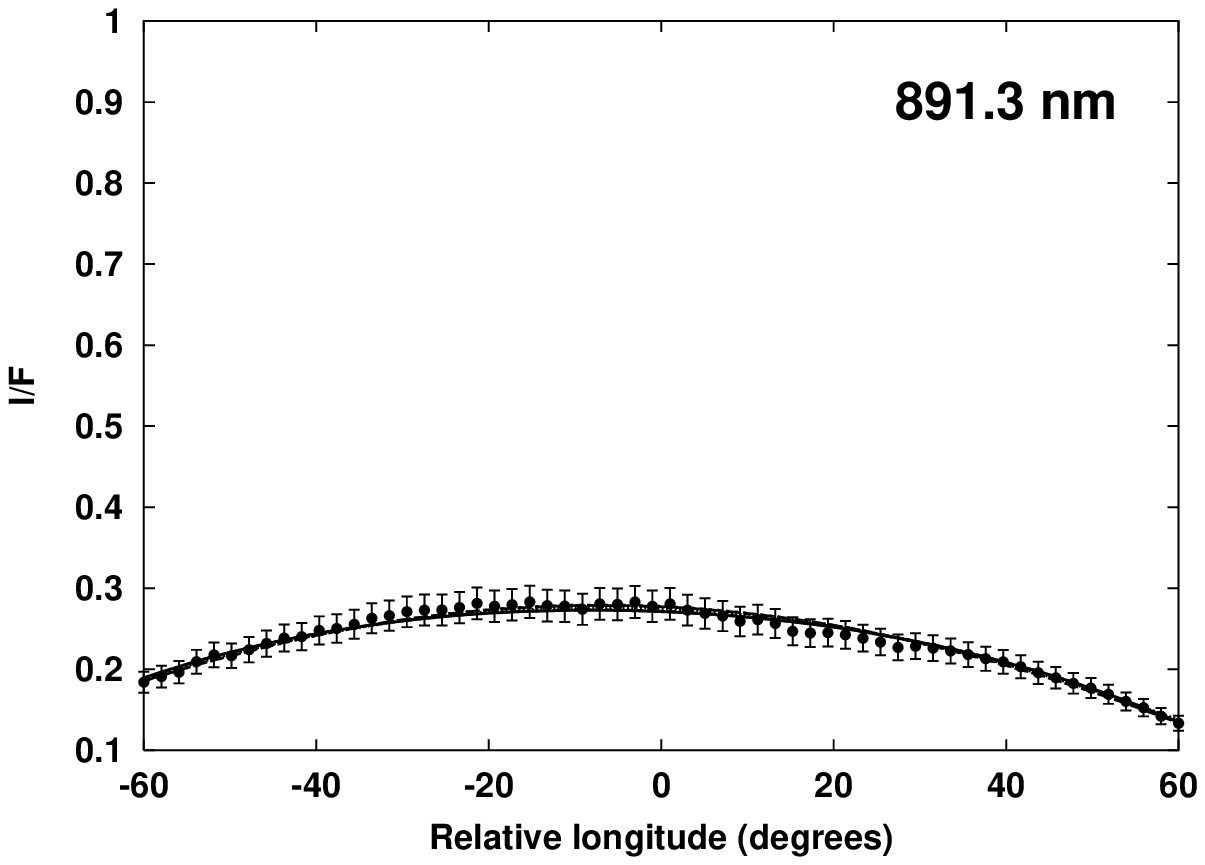}}
\end{center}
\end{minipage}&
\begin{minipage}{80mm}
\begin{center}
\scalebox{.7}{\includegraphics{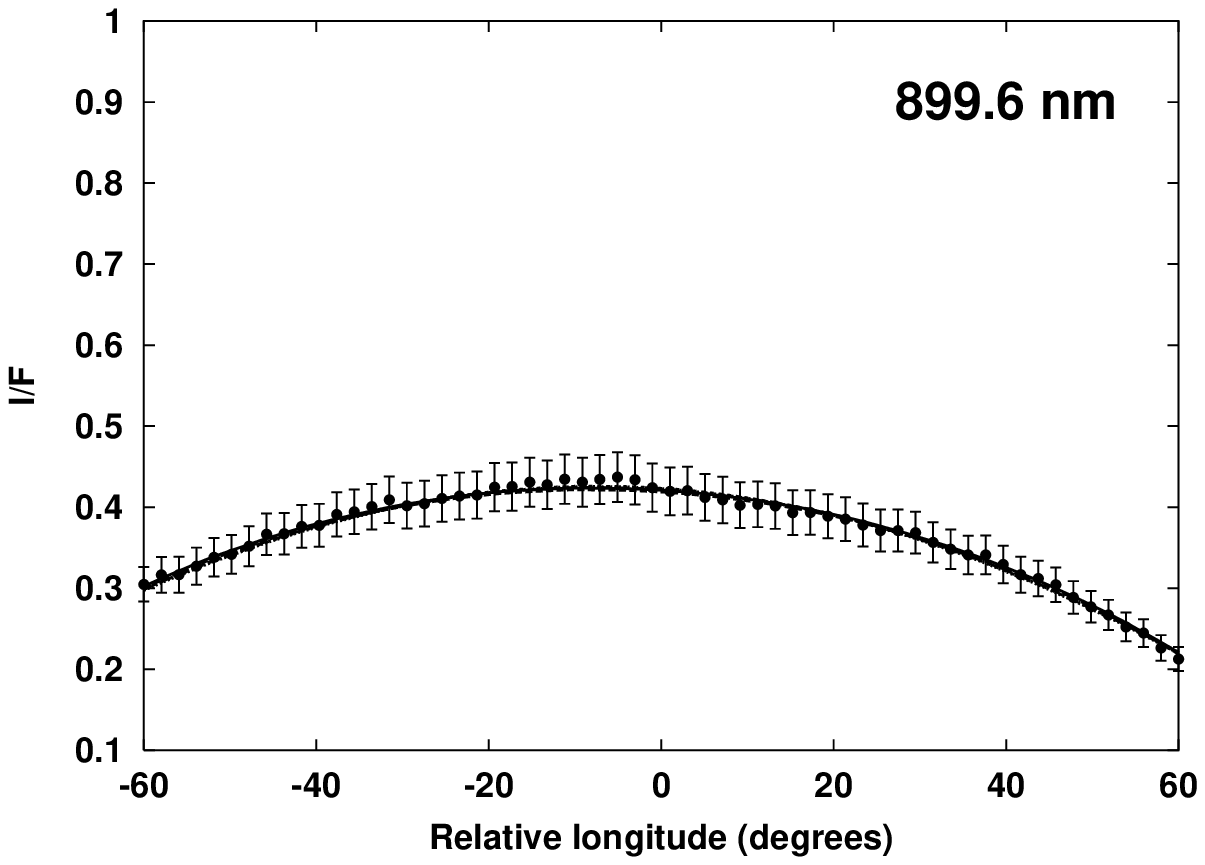}}
\end{center}
\end{minipage} \\
\begin{minipage}{80mm}
\begin{center}
\scalebox{.7}{\includegraphics{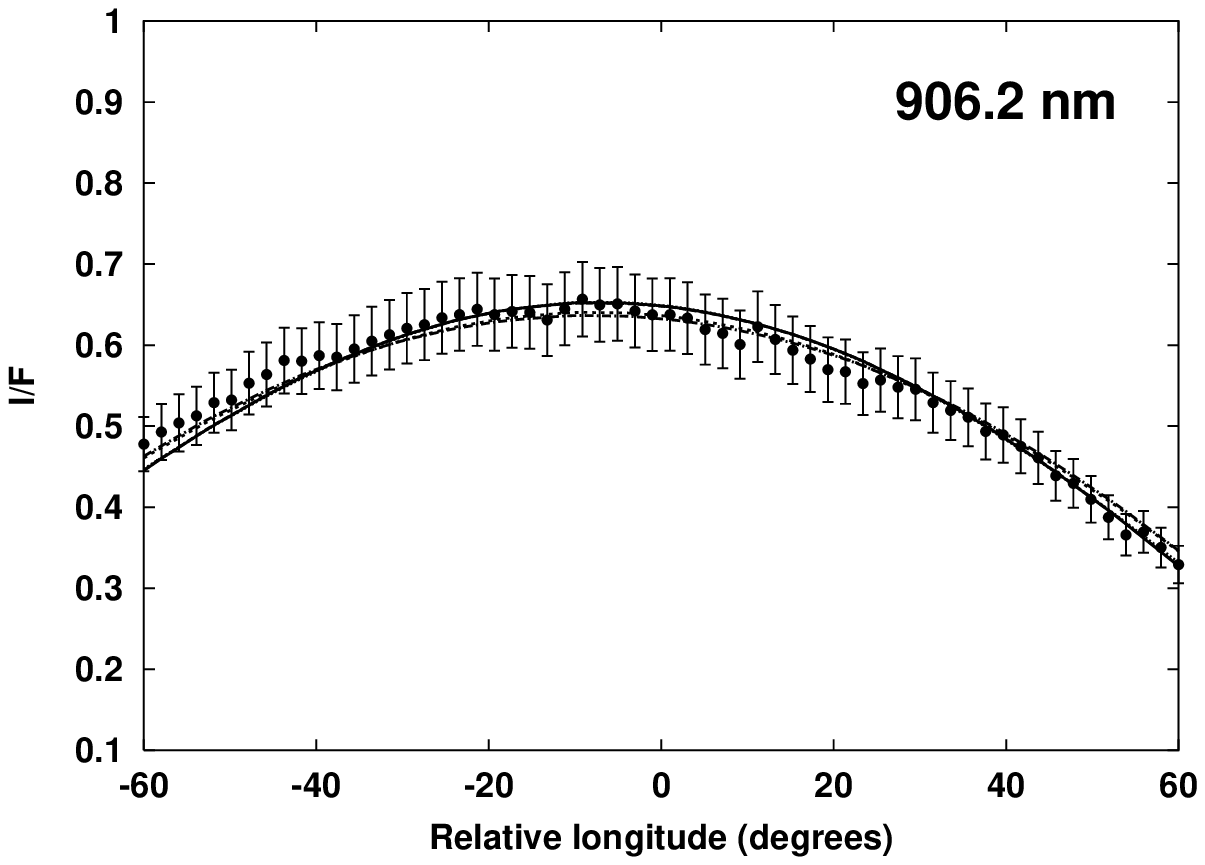}}
\end{center}
\end{minipage}&
\begin{minipage}{80mm}
\begin{center}
\scalebox{.7}{\includegraphics{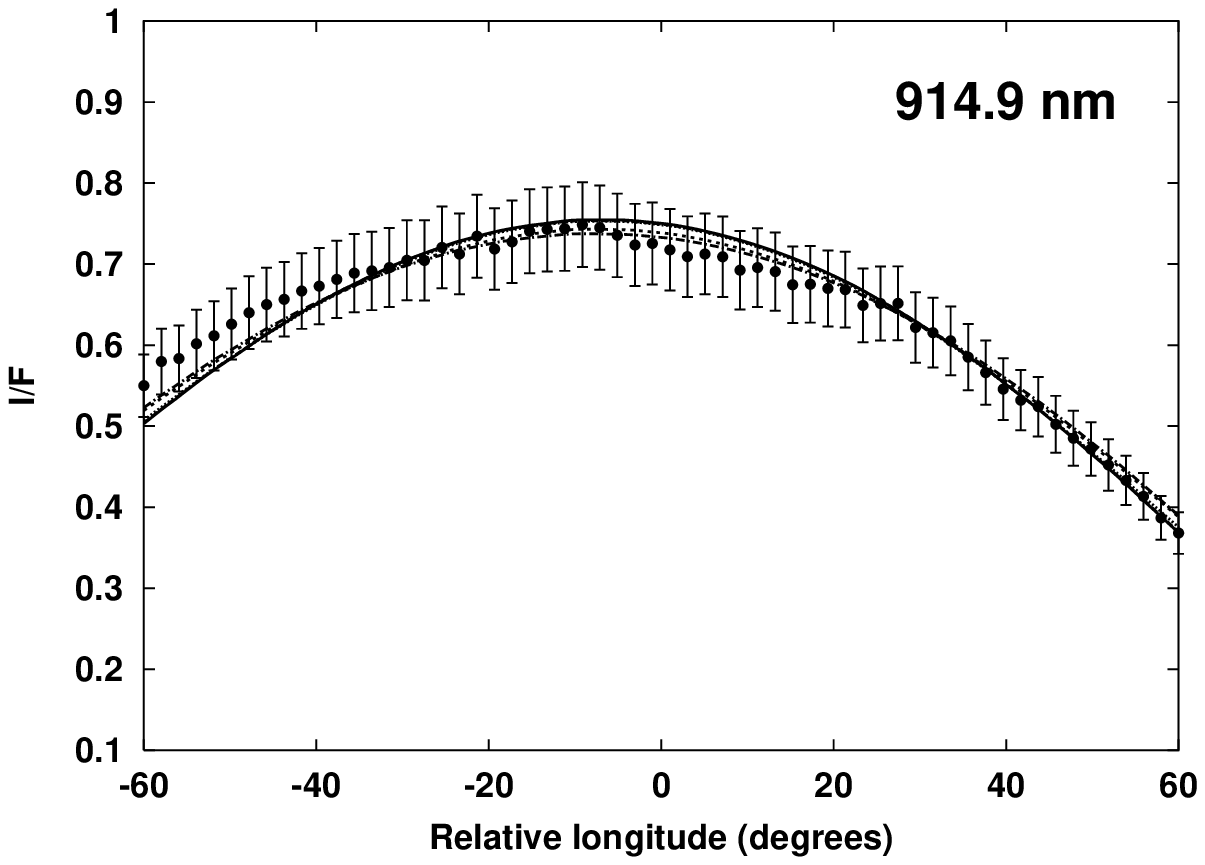}}
\end{center}
\end{minipage}
\end{tabular}

\begin{center}
\begin{minipage}{80mm}
\scalebox{.7}{\includegraphics{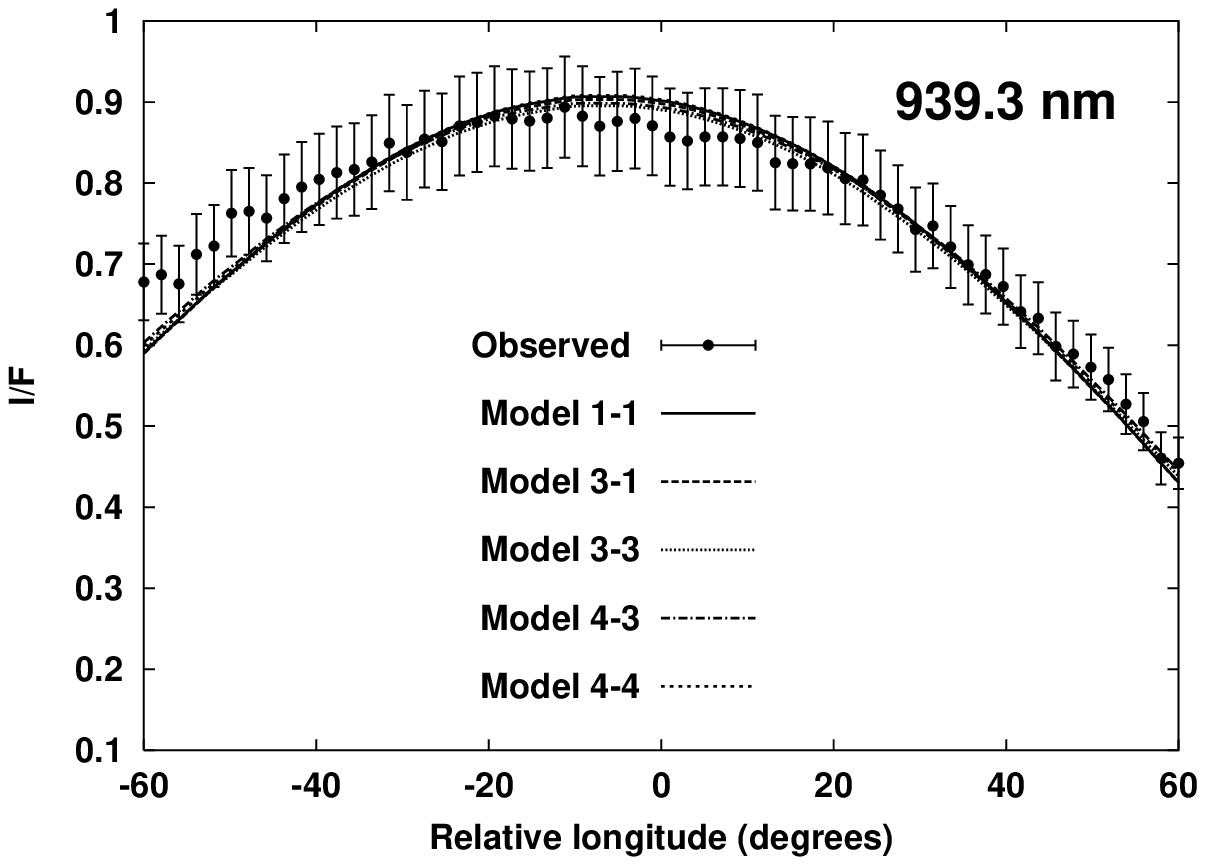}}
\end{minipage}
\end{center}
\caption{Graphical display of the best-fit results listed in Table 7 at
 five different wavelengths around 890 nm used in our modeling. The
 error bars show $\pm$7\% of calibration error. The solutions 3-1 (2)
 and 3-3 (2) are not plotted because they show no visually significant
 difference.}    
\end{center}
\end{figure}

\begin{figure}
\begin{center}
\begin{tabular}{cc}
\begin{minipage}{80mm}
\begin{center}
\scalebox{.7}{\includegraphics{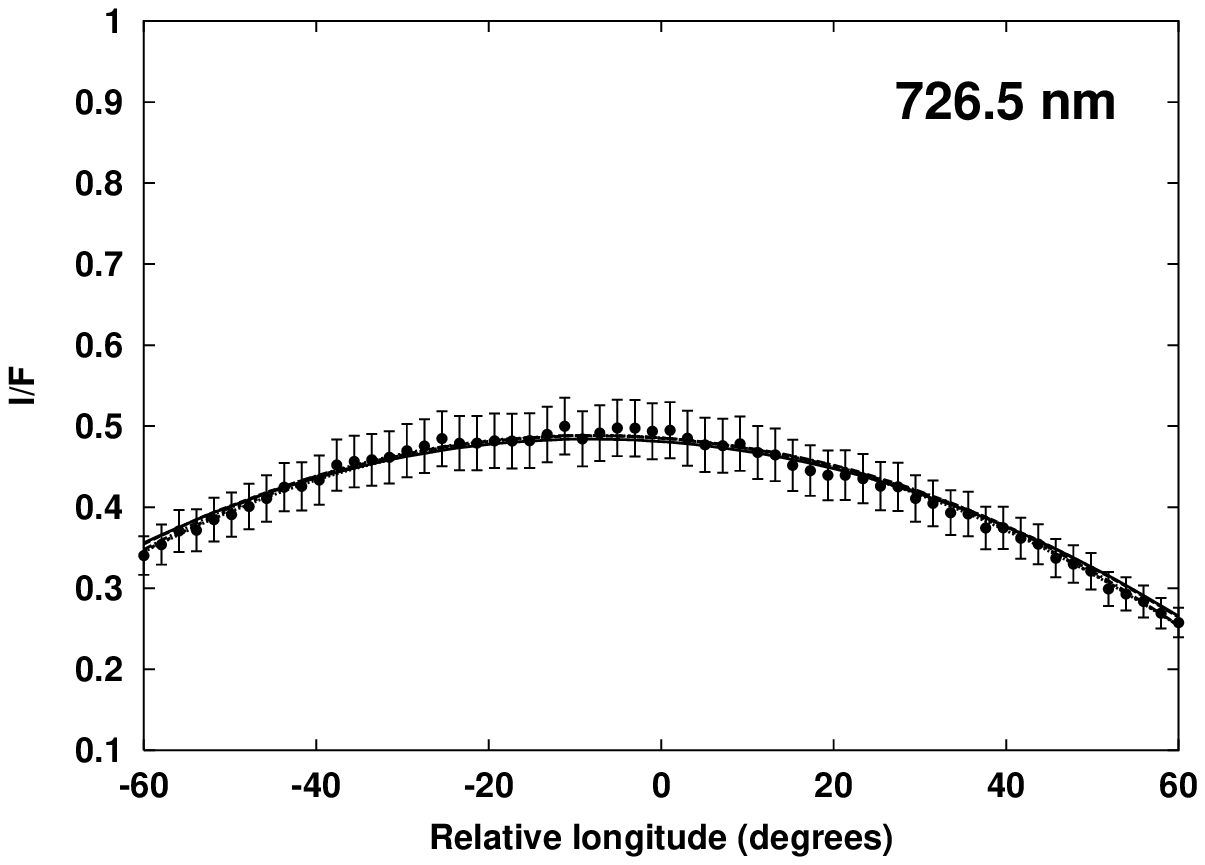}}
\end{center}
\end{minipage}&
\begin{minipage}{80mm}
\begin{center}
\scalebox{.7}{\includegraphics{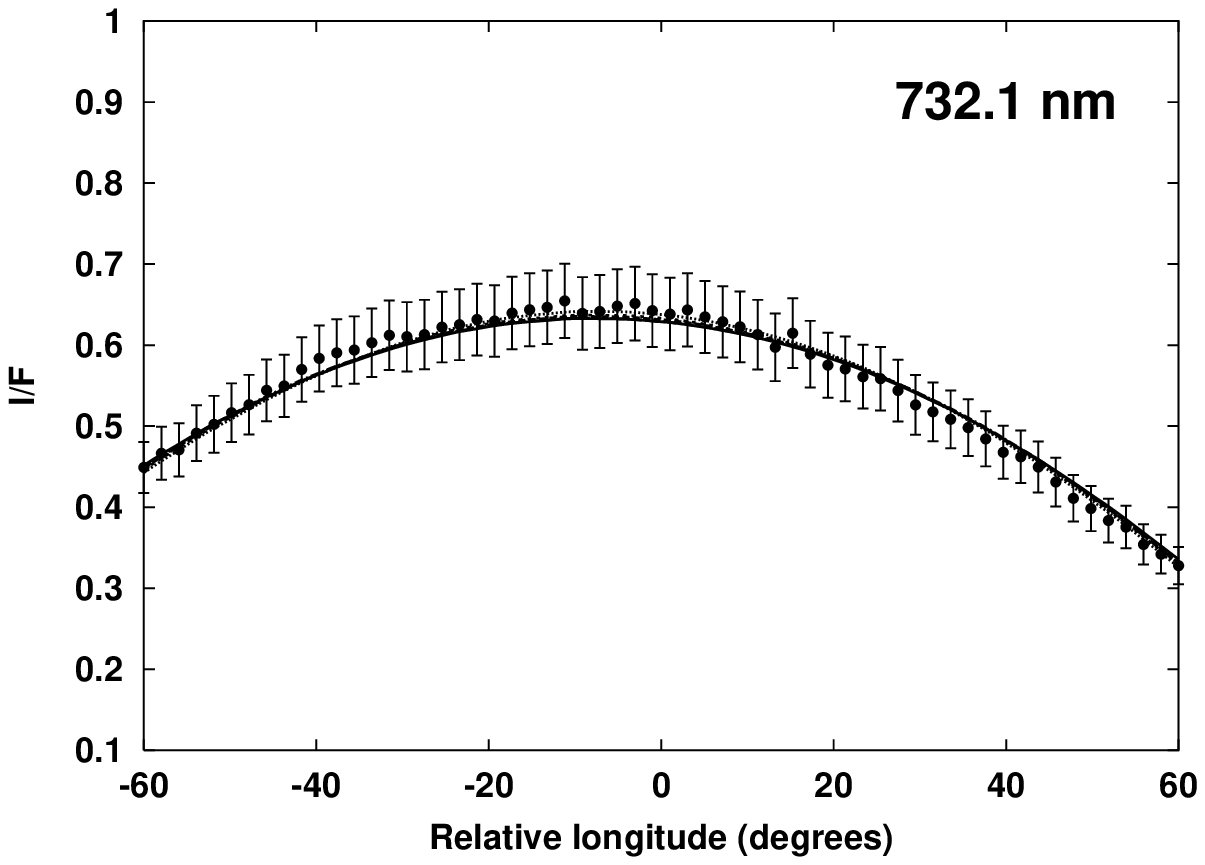}}
\end{center}
\end{minipage} \\
\begin{minipage}{80mm}
\begin{center}
\scalebox{.7}{\includegraphics{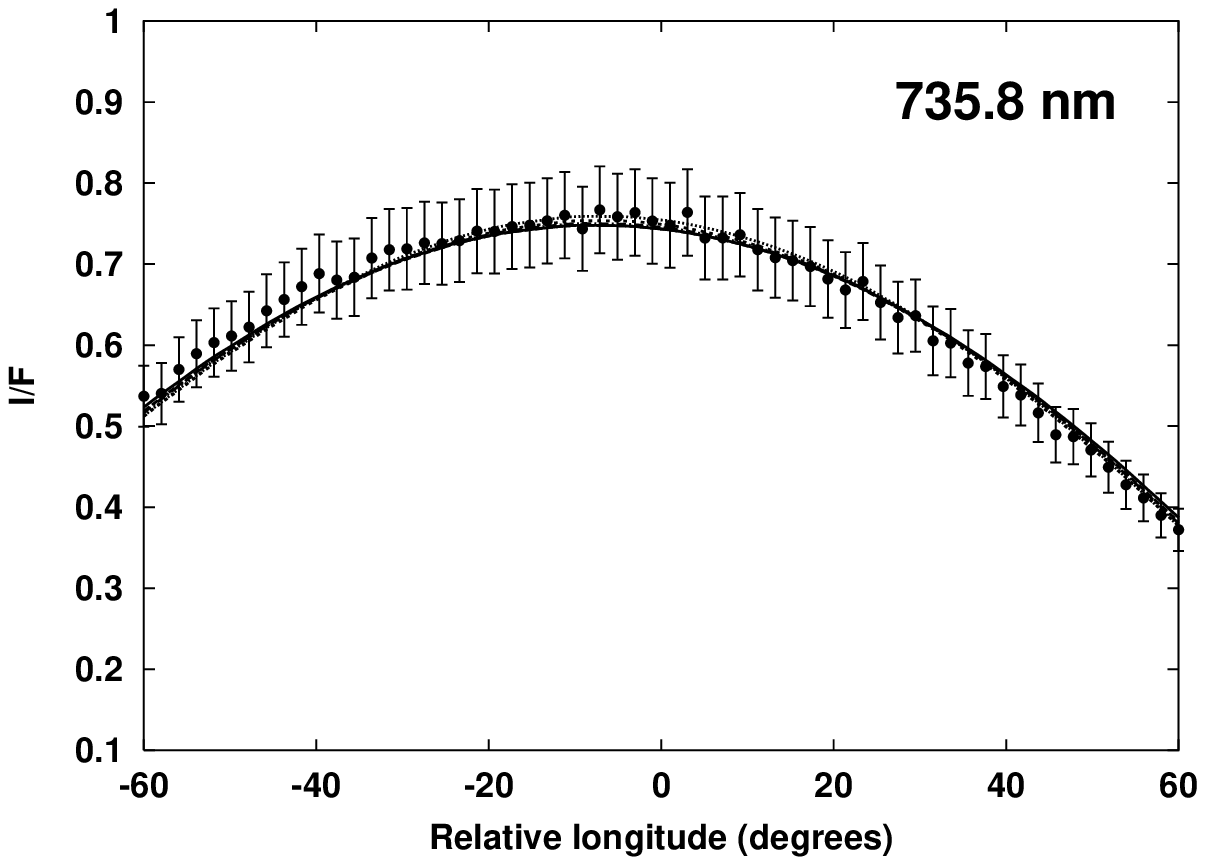}}
\end{center}
\end{minipage}&
\begin{minipage}{80mm}
\begin{center}
\scalebox{.7}{\includegraphics{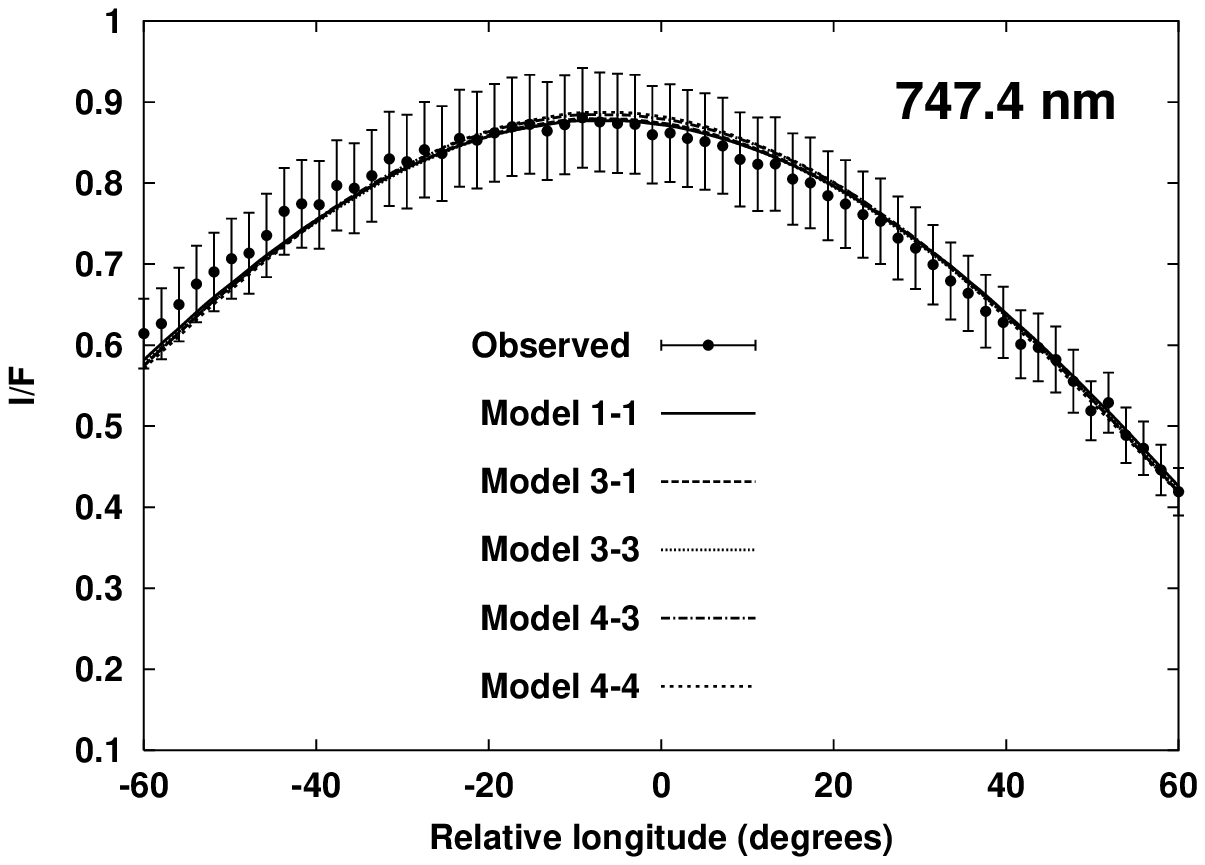}}
\end{center}
\end{minipage}
\end{tabular}
\caption{Same as Fig. 8 but at four different wavelengths around 727 nm.}    
\end{center}
\end{figure}

Judging from the average $\chi^2$ of the two data sets, we can state
that the model 4-3 solution is the best to reproduce the Saturnian
center-limb profiles of the equatorial region over the entire spectral
range in our observations. However, we note the following points : a)
there may be systematic biases associated with the selected cloud model
structures, b) the difference in $\chi^2$ between the model 4-3 solution
and the others is too small to be significant given our calibration 
uncertainty. The great difficulty in determining the best solution is
clearly seen when we plot the seven solutions together at all the nine
wavelengths (Figs. 8 and 9). Without any more objective measures to
distinguish one solution from another, we can not tell which solution
best represents the real cloud structure in Saturn's atmosphere. Therefore,
though we put more stress on the model 4-3 solution, we would rather
summarize some common features among the seven listed solutions than
take a risk of fully relying on the 'numerically-best' one. The drawback
of this approach is that the ensuing discussion inevitably tends to be
qualitative; nevertheless, the model-independent features among those
solutions will give us real physical insight about the equatorial cloud
structure of Saturn.           

\subsection{Result Overview}

First, we obtain somewhat better results from the two models with
infinite bottom cloud (4-3 and 4-4). This difference comes from the
better fit in the 939.3-nm profile. The large uncertainties in $DLP5$
imply that this continuum intensity is not strongly affected by the LCLD
altitude, as long as the LCLD top exists within the altitude range of
0.15 -- 1.5 bar. Therefore, we suggest that a region of higher aerosol
density located somewhere within 0.15 -- 1.5-bar level is preferred to
better reproduce the profile at 939.5 nm. Nonetheless, our calibration
uncertainty and model-dependent behaviors of the derived solutions
prevent us from determining if this high aerosol density region exists
as a separated cloud deck or a part of an extended cloud. These
model-dependent features are discussed in detail in the subsequent
section.          

We also note the narrow gap between the haze and UCLD ($DLP3$ is 10 -- 24
mb for all the haze-containing solutions). Since the haze bottom level
(= $DLP1+DLP2$ bar) always exists at 15 -- 30 mb, this suggests that the
tropospheric cloud in the Saturnian equatorial region extends up into
the stratosphere.     

Another noteworthy trend is the significantly smaller UCLD opacity in
 the 727-nm data fitting than in the 890-nm data fitting. The models
 without LCLD (1-1, 3-1 and 3-3) show a difference of roughly a factor
 of 2 in the UCLD optical thickness, and a difference of about 20\% is
 seen in the models with the LCLD (4-3 and 4-4). This opacity variation
 is required from the fact that the equatorial profiles near 727 nm are
 relatively dark given their relatively high brightness around 890
 nm. In fact, it is possible that the particle scattering efficiency
 becomes larger with longer wavelength over the spectral range of our
 data sets (700 -- 950 nm), provided the particle size is in an
 appropriate range. Figure 10 illustrates the wavelength dependence of
 the scattering efficiency with various effective aerosol radii and
 variances of Hansen's size distribution function.           

\begin{figure}
\begin{center}

\includegraphics{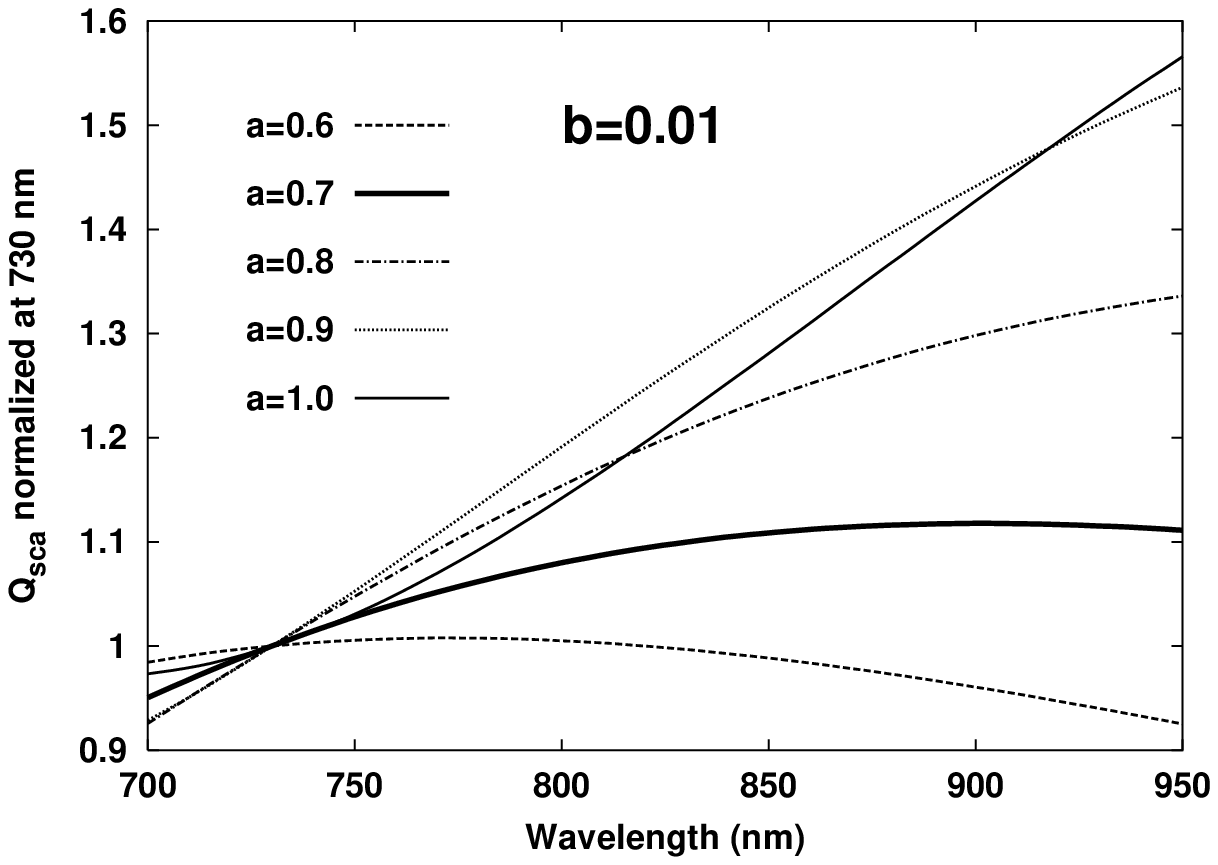}
\includegraphics{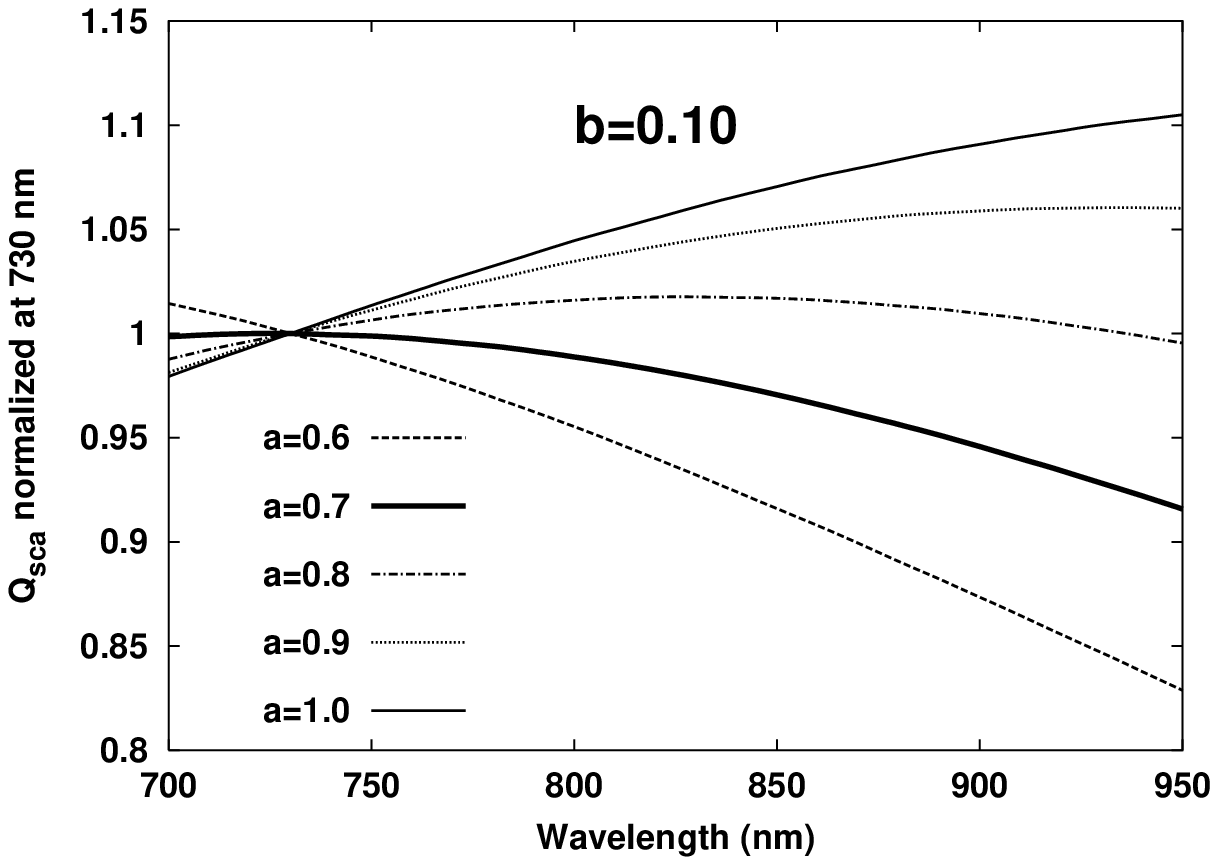}
\caption{Wavelength dependence of normalized scattering efficiency
 ($Q_{sca}$) over 700 -- 950 nm with different effective aerosol radii
 ($a$, in $\mu$m). $Q_{sca}$ is normalized at 730 nm, and
 $n_r=1.43$. Top: $b=0.01$. Bottom: $b=0.10$.}    
\end{center}
\end{figure}

The increase in aerosol opacity of the UCLD by roughly a
factor of 2 suggests a small variance in the aerosol size distribution
($b\sim0.01$) and an effective radius of about 0.9 $\mu$m. If we only
examine the opacity increase at longer wavelengths, we can still set a
lower limit of $a$ around 0.7 -- 0.8 $\mu$m, if $b \sim$ 0.01 -- 0.1 and
$n_r=1.43$.     

Regarding the haze parameters, we see an even more anomalous wavelength
 dependence. The haze optical thickness is disturbingly larger around
 890 nm than around 727 nm by roughly a factor of 4 or more, except for
 the model 3-1 (1) case. The large fractional errors in haze opacity and
 albedo at both 727 and 890 nm imply that the haze is relatively
 unimportant for the fitting results. Furthermore, the aerosol free
 layer between the haze and UCLD is always very thin in our
 solutions. Actually, in our additional trial simulations where we fixed
 $DLP3$ value to zero, we could obtain solutions of very similar cloud
 structures and fitting accuracies. This means that the thin
 aerosol-free layer between the haze and the UCLD is not essential for
 the profile fit. Based on these arguments, we suggest that the
 stratospheric haze is mostly mixed with the vertically extended
 tropospheric cloud. The apparent existence of separate haze layers in
 the haze-containing solutions are regarded as mere numerical
 artifacts. This inference is further supported by the great similarity
 between the solution 1-1 and solution 3-1 (1) in terms of the fitting
 accuracy, UCLD albedo, UCLD top and bottom altitudes and total opacity
 of the haze and UCLD. Recalling that both of these two solutions adopt
 the common 'H-G (red)' phase function in all the aerosol layers, we can
 say that those two results are practically equivalent.    

 A combination of 'Mie 0.15' phase function in the haze and 'H-G (red)'
 or 'Mie 1.5' in the UCLD tends to give slightly better fits to the
 data, as seen in models 3-3, 4-3 and 4-4. Following the above
 interpretation of the upper atmospheric structure, this supports the
 idea that the aerosol phase function at a lower altitude of the UCLD is
 more forward-scattering, indicating a larger aerosol size with
 increasing depth. In our view, the uppermost part of the UCLD is mainly
 composed of very small particles ($r_{eff} \sim$ 0.15 $\mu$m) probably
 including some haze components, and the lower main body of the UCLD
 consists of larger particles between 0.7 -- 0.8 and 1.5 $\mu$m.         

None of our cloud models, which are all longitudinally averaged, could
fit well the high dusk-side intensity near the two continuum wavelengths
({\it e.g.} 747.4 and 939.3 nm). The flat-fielding process is not likely
to be the origin of this feature because it does appear in the images
taken on the previous night (Feb. 7, 2002) at different west longitudes
despite the different positions and orientations on the detector, where
the flat-field profiles had totally different shapes along the
equatorial region. Another possibility, the image navigation error, is
also unlikely because the limb-fitting algorithm works better at
continuum wavelengths, where the contrast between the disk and
background is the highest. Therefore, the deviation from the
longitudinally averaged model in the dusk limb could represent a real
diurnal variation, though the deviation is not large enough compared
with our calibration errors. Any conclusive statements about this
phenomenon are reserved for future observations with higher photometric
accuracy.

\subsection{Dependence on the model structures and initial conditions}

Here we discuss some noticeable dependence of fitting results on assumed
cloud models and initial conditions for optimization process. Discussing
and summarizing these effects provides us with valuable lessons to be
kept in mind for the interpretation of the results presented herein as
well as in future modeling efforts. We found that the assumption of an
infinite bottom cloud and the choice of aerosol scattering phase
function both strongly influence the fitting results. We also found that
different initial settings for optimization can lead to very different
solutions.     

First, the total combined opacity of the haze and UCLD is
 systematically smaller if the LCLD exists. This is conceivable because
 the LCLD can account for some fraction of the required opacity to fit
 profiles in weak absorption bands and continua. Second,
 the physical thickness of the UCLD ($DLP4$) is much smaller in the
 models containing LCLD. This trend is related to the aforementioned first
 feature. To maintain a certain reflectivity in an absorption band with
 a small total aerosol opacity in upper aerosol layers, those aerosols
 needs to be concentrated in a narrow pressure range, resulting in a
 physically thin cloud. This is the main cause of the surprisingly
 high altitudes of the UCLD bottom in our 4-3 and 4-4 solutions, which
 unrealistically indicate that the whole UCLD ($\tau = 5-9$) exists
 above the tropopause (Saturn's tropopause is located around the 0.1-bar  
 level). Therefore, we consider this high UCLD bottom level in those 
 two solutions as an numerical 'illusion' associated with the assumption
 of the LCLD, and conclude that a cloud in the troposphere extends into
 the stratosphere. Third, the gaps between aerosol layers are larger in
 the models with the LCLD. This is probably because the UCLD is more
 narrowly confined and we do not consider any gas absorption in the
 LCLD. Namely, the increase in gas absorption in thicker aerosol-free
 layers compensates for reduced absorption in physically thin clouds and
 LCLD. If we allow for mixing of gas in the bottom cloud, the $DLP5$
 value will become smaller and the vertical structure will approach
 that of the solutions 3-1 or 3-3. From these discussions, we infer that
 the adoption of an infinite bottom cloud in a cloud model introduces:
 high altitude, low opacity and a small physical thickness of an upper
 aerosol layer, and also introduces relatively large gaps between
 aerosol-containing layers.          

Regarding the effect of phase function, it is evident that the opacity
of the UCLD is strongly influenced by the chosen type of scattering
phase function when we compare the solutions of model 4-3 and
4-4. Generally, more forward-scattering aerosols tend to require more
opacity to reproduce observed intensities in nearly
back-scattering direction, {\it i.e.} observations at small phase angles.      

The dependence on initial conditions is clear in the comparison
between the solutions 3-1 (1) and 3-1 (2), and to a lesser extent
between 3-3 (1) and 3-3 (2). These examples demonstrate that different
initial conditions for optimization can result in very different
solutions of equally good fit through simultaneous adjustment of
multiple variables. What is worse, we found no apparent correlation
between the initial conditions and corresponding final solutions in our
analysis.  

Therefore, when a forward-modeling approach is taken to analyze an
atmosphere, one must : a) fix as many atmospheric parameters as possible 
based on reasonable assumptions or observational facts, b) be very
careful about the validity of the assumptions about cloud structure
model, c) try as wide a range of initial conditions as possible, and d)
take great care in interpreting modeling results, recognizing the
influence of the assumptions on the results.

\subsection{Comparison with previous results}

In this section, we compare our results with previous studies. Despite
the uniqueness of our modeling method and difference in data acquisition
epoch, we see some interesting similarities and differences.   

In the equatorial region of Saturn, Stam {\it et al.} (2001) found the
stratospheric haze bottom around 20 mb level, the UCLD top around 60 mb
level and the tropospheric cloud bottom around 300 mb level by inverting
the near-IR spectra of Saturn taken in August, 1995. The haze bottom
level is consistent with our results ($P = 14-33$ mb). Their UCLD top is
somewhat lower than ours ($P = 24-57$ mb), and the tropospheric cloud
bottom is in rough agreement with our solutions without LCLD within the
errors. The important difference is that Stam {\it et al.} (2001) argue
for distinctly separated aerosol layers, in contrast to our preference
for a diffuse, extended UCLD. This could be a real time-variation in the
equatorial cloud structure related to seasonal changes over six and a
half years, corresponding to roughly a quarter of a Saturnian year.     

Acarreta and S\'anchez-Lavega (1999) present three sets of modeling results
based on the observations of Saturn's equatorial region before, during,
and after a convective outburst in 1990 (S\'anchez-Lavega {\it et al.}
1994). Although Acarreta and S\'anchez-Lavega (1999) assumed an infinite
cloud at the bottom in their model, the large tropospheric opacity and
low aerosol albedo in their results screened out most of the effect from
the bottom cloud. Consequently, their results show more similarities to our
solutions without LCLD than to the solutions with LCLD. The result of
their data set A, which corresponds to the beginning of the disturbance,
is not similar to any of our solutions. The albedo and bottom level of
the UCLD are much lower than ours. Though their tropospheric opacity of
30 appears close to the values in our solutions 3-1 (2), 3-3 (1) and 3-3
(2), their UCLD is much more vertically extended and therefore the
opacity per unit pressure range is much smaller. These differences
result from the significantly lower brightness of Saturn at the epoch of
their observation. Crudely estimating from their geometrically corrected
$I/F$ at latitude $5^{\circ}$ (S\'anchez-Lavega {\it et al.} 1994), the
equatorial region of Saturn at the time of their set A observation
(October 1990) was darker than our data by more than 30\% and 50\%
around 727 and 890 nm, respectively. These deviations are much larger
than their photometric error (10--15\%) and ours (5\%). The Great White
Spot core analysis of Acarreta and S\'anchez-Lavega (1999) (data set O
in their paper) shows very high aerosol albedos and high UCLD top and
bottom levels  similar to our model 1-1 and 3-1 (1) results. Their data
set B result, which is of the mature phase of the disturbance, also
resembles our solutions of 1-1 and 3-1 (1) in terms of the opacity,
albedo and bottom altitude of the UCLD. In addition, the wavelength
dependence of the UCLD opacity ($\tau \sim 9$ around 727 nm and $\tau
\sim 12$ around 890 nm) is consistent with our findings. In their set C
result (fully evolved phase of the disturbance), their UCLD becomes
thicker and more extended, and the aerosol albedo significantly lowers
again. The cloud structure shows similarities to our model 3-3 (2)
result with respect to the UCLD top and bottom levels, but with much
smaller opacity per unit pressure range. 
 
Our model 4-4 has basically the same cloud structure and phase functions
as those in Ortiz {\it et al.} (1996). In addition, the high
reflectivity in the equatorial region over 700-950 nm in our observation
most resembles their August 1992 and May 1993 data sets (Ortiz {\it et al.}
1995). Accordingly, we compare our model 4-4 solution with their
results of equatorial region (Latitude $0^{\circ}$) for August 1992 and
May 1993. The agreement in the haze top level is very good. Their much
larger UCLD opacity ($\sim 30-100$) and much deeper UCLD bottom ($P
\sim0.2-1.0$ bar) are likely to result from their very low LCLD level
($P$=1.8 bar). Since the LCLD altitude is fixed at a much lower level in
their model, a significantly larger opacity is necessary in the upper
troposphere to obtain the required reflectivity in absorption bands,
while it must be vertically stretched to account for the appropriate
amount of gas absorption at continuum wavelengths. As a result, the LCLD
has only a minor effect on the result, and their upper cloud
structure becomes rather similar to our solutions with no LCLD ({\it
e.g.} 3-1 (2), 3-3 (1) and 3-3 (2)). In fact, the haze top level, haze
opacity, UCLD opacity and UCLD bottom level in the 'supersigma'
results of August 1992 of Ortiz {\it et al.} (1996) agree well with our
solution of 3-3 (2). However, in contrast to these similarities, their
albedo values are systematically lower than ours. This is probably
because the calibrated intensity in our data is systematically higher
than their brightest August 1992 data by $\sim$10\% around 750 nm and
$\sim$40\% around 890 nm (Ortiz {\it et al.} 1995). Considering their
photometric error ($\sim$10\%) and ours ($\sim$ 5\%), the discrepancy at
750 nm is still barely within the uncertainty, although the intensity
difference at 890 nm is outside the error bars. 

The HST data analysis by Karkoschka and Tomasko (1993) reports a much
smaller tropospheric opacity than ours around the equator ($\tau \sim 4$
at 340 nm) mainly because their observation was done in July 1991, when
Saturn was much darker at 890 nm. In spite of that, their conclusions
are qualitatively consistent with ours in that the tropospheric cloud in
the equatorial region may extend into the stratosphere and that the
aerosol size decreases with increasing height. The high albedos of the 
stratospheric haze and tropospheric aerosol at 890 nm is also quite
consistent with our solutions.    
  
In the study of the optical spectrum of the Saturnian equator taken in
the late 1980's, Karkoschka and Tomasko (1992) obtained a thick
tropospheric cloud ($\tau \sim 30$) stretching into the stratosphere
($\sim 50$ mb). Due to the large opacity in the upper troposphere, their
infinite lower cloud should have minor contribution to the outgoing
radiation. Moreover, they assumed an average TTHGF of Tomasko and Doose
(1984). Accordingly, it is quite natural that their best-fit model
structure qualitatively resembles our results without the LCLD. Their
vertically extended tropospheric cloud structure and their albedo values
agree well with our solutions without LCLD (3-1 (2), 3-3 (1) and 3-3
(2)). Their UCLD opacity of 30 down to the 1.8-bar level over 600--800
nm is much smaller than our values around 727 nm ($\sim$20 down to 0.5
bar) when scaled to the same pressure range. This is again because the
Saturnian intensity was much lower at the epoch of their observation
(roughly estimating from their equatorial albedo data, the peak values
of $I/F \leq \sim$ 0.35 and $\leq \sim$ 0.15 at 727 and 890 nm,
respectively).    

West (1983) tried to fit his ground-based data (West et al. 1982) in the
619-, 725- and 890-nm methane bands. His cloud model compares with
our model 1 structure, simply consisting of an aerosol free layer and a
diffuse infinite cloud below that. However, the significantly lower
reflectivity of his data, especially around 899.6 nm, resulted in lower 
albedo values (0.992 and 0.991 at 750 and 936.5 nm, respectively) and
much smaller aerosol opacity per unit pressure range ($\tau \sim 5$ over
180--700 mb) in the equatorial region ($-6$ to $-11^{\circ}$) than our
solution 1-1.      

As we see, most of the previous observations show significantly lower
intensities than ours over 700--950 nm. Therefore, we conclude that a
significant brightening of the equatorial region of Saturn occurred over
the last decade. This increase in the observed brightness is mainly
interpreted either as the increase in aerosol albedo or the increase in
the upper cloud opacity per unit pressure range, or both. The
only result that simultaneously showed similar opacities, albedos and
cloud altitudes to our solutions is those of the data set O and B in
Acarreta and S\'anchez-Lavega (1999). In particular, we note the simultaneous
emergence of significant increase in aerosol albedo and the wavelength
dependence of aerosol opacity during the 1990 storm. Hence, we suggest
that the equatorial aerosols in February 2002 had similar size and
composition to those of the 1990 disturbance. It seems that fresh
materials from the deep Saturnian atmosphere had been gradually
transported to the upper atmosphere to form reflective aerosols over
the last decade, not as in the sudden convective event in 1990. This may
be a manifestation of a slow seasonal change in the upper Saturnian
atmosphere.

\section{Conclusions}

We draw the following conclusions from our modeling analysis of the
Saturnian equatorial region and the comparison with previously published
results.    

\begin{itemize}
\item The assumption of an infinite bottom cloud does improve the
      accuracy of the center-limb profile fit in the equatorial region
      of Saturn, though good fitting accuracy can be obtained without
      the infinite cloud given our 7\% of calibration error.    
\item A cloud model having higher aerosol density in the lower
      troposphere (0.15 -- 1.5 bar) is preferred, but we could not
      distinguish whether this high density region exists as a separated
      layer or a part of a diffuse cloud.
\item The tropospheric cloud extends into the stratosphere (higher than
      0.1-bar level) in the equatorial region of Saturn. This could
      suggest existence of an upward atmospheric motion on a planetary
      scale in Saturn's equatorial region.    
\item The wavelength dependence of the UCLD opacity
      suggests a lower limit of average aerosol size around
      0.7--0.8 $\mu$m if the material is ammonia ice, provided the
      effective variance of the size distribution is 0.01--0.1.   
\item The average aerosol size of the extended upper tropospheric cloud
      increases with depth from about 0.15 $\mu$m in the stratosphere to
      between 0.7--0.8 and 1.5 $\mu$m in the troposphere if $n_r=1.43$.  
\item The aerosol property in February 2002 is similar to that seen
      during the 1990 equatorial disturbance. This may be a sign of
      long-term dredging up of fresh materials associated with a
      seasonal change in Saturn's upper atmosphere.          
\end{itemize}

Our most preferred cloud model that bears all the above described
features is illustrated in Fig. 11.

\begin{figure}
\begin{center}
\includegraphics{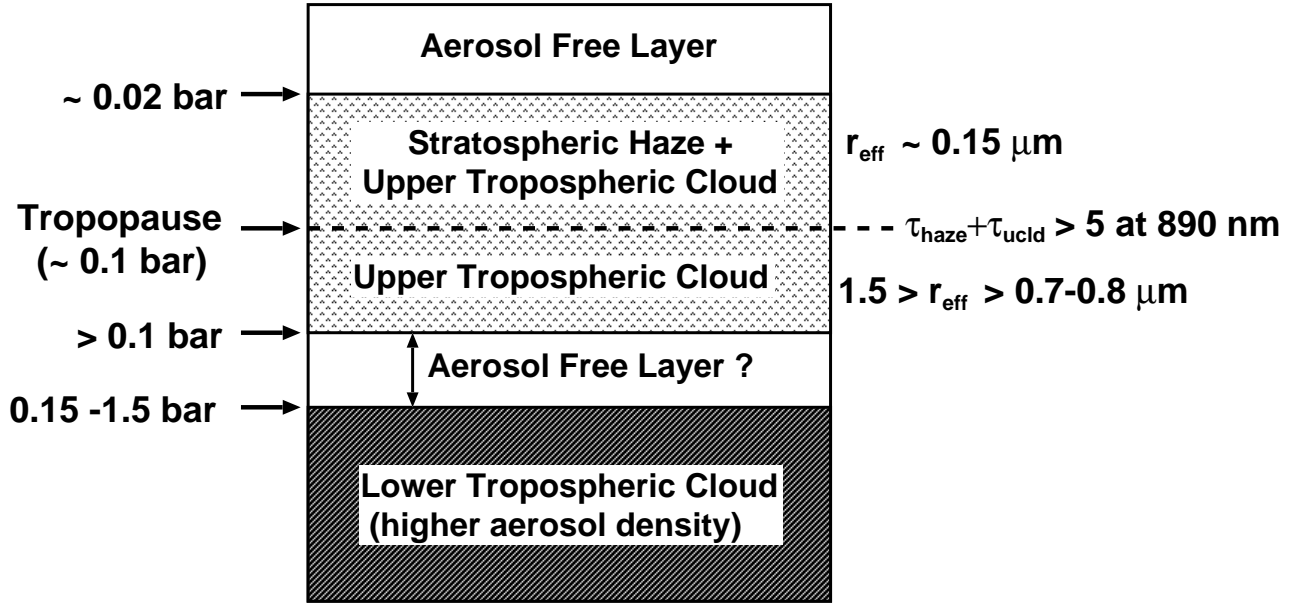}
\caption{The most favorable model in our modeling analysis of the
 equatorial region of Saturn. }    
\end{center}
\end{figure}

We discovered that the best-fit modeling parameters depend heavily on
assumed cloud structures, choice of aerosol scattering functions and
initial optimization conditions. We are presently applying the same
fitting method to other latitudes in Saturn's southern hemisphere and
comparing them to discern any latitudinal variation in Saturn's
atmospheric structure. We will also examine the effect of choices of
atmospheric composition on the modeling results. These results will be
presented in a forthcoming publication.      

The Cassini spacecraft will arrive at the Saturnian system in late
2004. It will undoubtedly provide a great deal of new data covering a
wide range of phase angle and of very high spatial resolution. Those
unique data, which can never be obtained from the earth, will greatly
help to examine the Saturn's aerosol properties and local meteorological
phenomena. Nevertheless, ground-based observations with an AOTF camera
offer some advantages over the observations by Cassini, such as the
extended temporal coverage to track down the seasonal change in Saturn's
atmosphere and the spectral agility that is not obtainable with
Cassini's limited number of fixed narrow-band filters (619-, 727- and
890-nm methane bands and adjacent continuum wavelengths). Therefore,
data sets acquired by Cassini and an AOTF camera will be nicely
complementary to each other, and the combination of the analysis results
from those two different data sets will elucidate Saturn's cloud
structure and its variability over time.        
\vspace{1cm}

\centerline{\bf\Huge Appendix }
\vspace{5mm}

\centerline{\bf\Large A: Photometric Reduction of Saturn Data}
\vspace{5mm}

The intensity of light from Saturn represents the fraction of incident
solar radiation at Saturn that is scattered by its
atmosphere. Accordingly, it is convenient to calibrate Saturn's
intensity in terms of the incident solar flux, $\pi F_{\odot}$. We
finally obtain the quantity $I/F_{\odot}$, which is a modified ratio
of the Saturnian intensity $I$ and the incident solar flux $\pi
F_{\odot}$. All the arguments below apply to a certain wavelength
($\lambda$) of interest.  

 First, we seek the formula for the Saturnian intensity. We focus
 on a single pixel on the Saturnian image and express the intensity from that
 pixel as $I$. Then, the total radiation energy, $E$, arriving at Earth
 per unit time from that pixel is expressed as

\begin{eqnarray} 
E = I \cos \theta \cdot dS \cdot d\omega_E \nonumber \\ 
  = I \cos \theta \cdot dS \cdot \displaystyle{\frac{S_E}{\Delta^2}}, 
\end{eqnarray}

where
 
\begin{description}
\item[$\theta$:] The zenith angle of the Earth seen at the considered pixel on Saturn,
\item[$dS$:] The actual area covered by the pixel on Saturn, 
\item[$d\omega_E$:] The solid angle subtended by the Earth seen
	   from the surface of Saturn,
\item[$S_E$:] The cross-section of the Earth when projected in the
	   direction of Saturn, 
\item[$\Delta$:] The distance between the Earth and Saturn.
\end{description}
\vspace{5mm}

 Therefore, the flux $F$ falling on the Earth from that pixel is  

\begin{equation}
F = \frac{E}{S_E} = I \cdot \frac{cos \theta dS}{\Delta^2}. 
\end{equation}

Note that $\displaystyle{\frac{cos \theta dS}{\Delta^2}}$ is the
solid angle subtended by the pixel seen from the Earth; we define this
as $\omega$. Then,

\begin{equation}
I = \frac{F}{\omega}. \label{aa}
\end{equation}

Next, if we define the solar flux incident on Saturn as $\pi
F_{\odot,S}$ and that on the Earth as $F_{\odot,E}$ (here, we deliberately
define the Solar flux on the Earth as $F_{\odot,E}$, not as $\pi
F_{\odot,E}$,  for later convenience), 

\begin{equation}
\pi F_{\odot,S} = \frac{1}{d_s^2} \cdot F_{\odot,E} \label{bb}
\end{equation}

where $d_s$ denotes the heliocentric distance of Saturn. 

If we compare the solar flux with the stellar flux of the designated
comparison star,  

\begin{equation}
m_{star}-m_{\odot} = -2.5 \log_{10}\left(\frac{F_{star}}{F_{\odot,E}}\right),
\end{equation}

where
 
\begin{description}
\item[$m_{star}$:] Magnitude of the comparison star,
\item[$m_{\odot}$:] Magnitude of the Sun,
\item[$F_{star}$:] Stellar flux,
\item[$F_{\odot,E}$:] Solar flux incident on the Earth. 
\end{description}

Therefore, 

\begin{equation}
F_{\odot,E}= F_{star} \cdot 10^{0.4(m_{star}-m_{\odot})}. \label{cc}
\end{equation}

From (\ref{bb}) and (\ref{cc}), we obtain

\begin{equation}
F_{\odot,S}= \frac{1}{\pi d_s^2} F_{star} \cdot 10^{0.4(m_{star}-m_{\odot})} \label{dd}
\end{equation}

As a consequence of (\ref{aa}) and (\ref{dd}), we obtain the following
formula.  

\begin{eqnarray}
\frac{I}{F_{\odot,S}}= \frac{\pi d_s^2 }{\omega} \cdot \frac{F}{F_{star}} \cdot 10^{0.4(m_{\odot}-m_{star})} \nonumber \\ 
= \frac{\pi d_s^2 }{\omega} \cdot 10^{0.4(m_{\odot}-m_{star})} \cdot \frac{F}{F_{star2}} \cdot \frac{F_{star2}}{F_{star}} \nonumber \\
= \frac{\pi d_s^2 }{\omega} \cdot 10^{0.4(m_{\odot}-m_{star})} \cdot \frac{DN}{DN_{star2}} \cdot \frac{DN_{star2}}{DN_{star}}
\end{eqnarray}

where, 

\begin{description}
\item[$DN$:] Observed digital number on the detector obtained at the
	   considered pixel on Saturn image,
\item[$DN_{star}$:] Observed total digital number of the primary standard
	   star on the detector,
\item[$F_{star2}$:] Flux from the secondary standard star,
\item[$DN_{star2}$:] Observed total digital number of the secondary standard
	   star on the detector. 
\end{description}

This last equation means that we need only the following
information to calibrate Saturnian images : heliocentric
distance of Saturn at the time of observation ($d_s$), angular pixel
scale on the detector ($\omega$), magnitudes of the Sun and the primary
standard star ($m_{\odot}$ and $m_{star}$, respectively) and the digital
numbers on the detector obtained from the observations of Saturn,
primary and secondary standard stars. 
\vspace{1cm}

{\bf\Large B: Calculation of opacity and albedo in an atmospheric layer} 
\vspace{5mm}

In our model, the total opacity ($\tau_{total}$) and effective
single-scattering albedo ($\varpi_{0eff}$) of each given atmospheric
layer is computed as follows. First, given the pressure levels of layer
boundaries, the opacity by Rayleigh scattering gas and its albedo are
given as   

\begin{equation} 
\tau_R = \tau_{R,s} + \tau_{R,a} + \tau_{\mbox{\tiny $CH_4$}}
\end{equation}

and

\begin{equation} 
\varpi_{0R} = \displaystyle{\frac{\tau_{R,s}}{\tau_R}}
\end{equation}

where

\begin{description}
 \item[$\tau_{R,s}$:] Scattering opacity by Rayleigh gas, 
 \item[$\tau_{R,a}$:] Absorption opacity by Rayleigh gas in continua, 
 \item[$\tau_{\mbox{\tiny $CH_4$}}$:] Opacity by methane absorption. 
\end{description}

Here,

\begin{equation} 
\tau_{R,s} =  \displaystyle{\frac{\Delta p}{\overline{m} g} \sum_{i}} \alpha_i k_{i,s}
\end{equation}

\begin{equation} 
k_{i,s} = \displaystyle{\frac{32 \pi^3}{3} \cdot \frac{1}{N^2} \cdot \frac{(n_{\lambda,i}-1)^2}{\lambda^4} \cdot \left( \frac{6+3\delta_i}{6-7\delta_i}\right)} 
\end{equation}

\begin{equation} 
n_{\lambda,i}-1=A_i(1+\displaystyle{\frac{B_i}{\lambda [\mu m]^2}})
\end{equation}

\begin{description}
 \item[$\alpha_i$:] Mixing ratio of {\it i} th atmospheric component 
 \item[$k_{i,s}$:] Scattering coefficient of {\it i} th atmospheric component 
 \item[$\Delta p$:] Pressure difference between top and bottom of the layer 
 \item[$\overline{m}$:] Mean molecular weight 
 \item[$g$:] Gravity acceleration
 \item[$N$:] Loschmidt number $\equiv$ Number of atoms or molecules in
	    unit volume under the standard temperature and pressure
	    (hereafter, S.T.P. : T=$273.15$[K], P=$1.01325 \times 10^5$[Pa])  
 \item[$\delta_i$:] Constant to obtain the depolarization factor$\left(\equiv \displaystyle{\frac{6+3\delta_i}{6-7\delta_i}}\right)$ of {\it i} th atmospheric component (Cox 2000)
 \item[$n_{\lambda,i}$:] Refractive index of {\it i} th atmospheric component
	    at wavelength $\lambda$ under the S.T.P. 
 \item[{$\lambda [\mu m]$}:] Wavelength in units of $\mu m$
 \item[$A_i, B_i$:] Constants to compute the refractive index of {\it i}
	    th atmospheric component at $\lambda [\mu m]$ (Cox 2000)
\end{description}
\vspace{1cm}

$\tau_{R,a}$ is obtained when the continuum albedo of Rayleigh gas,
$\varpi_{0R,c}$, is given as a free or fixed parameter. 

\begin{equation} 
\tau_{R,a} = \tau_{R,s} (\displaystyle{\frac{1}{\varpi_{0R,c}} -1})
\end{equation} 

where, 

\begin{equation} 
\varpi_{0R,c} \equiv \displaystyle{\frac{\tau_{R,s}}{\tau_{R,s} + \tau_{R,a}}}
\end{equation}

 In the methane bands, the opacity by methane absorption is calculated as

\begin{equation} 
\tau_{\mbox{\tiny $CH_4$}} =  \displaystyle{\frac{\Delta p}{\overline{m} g}} \alpha_{\mbox{\tiny $CH_4$}} k_{\mbox{\tiny $CH_4$}}
\end{equation}

\begin{description}
 \item[$\alpha_{\mbox{\tiny $CH_4$}}$:] Methane mixing ratio
 \item[$k_{\mbox{\tiny $CH_4$}}$:] Methane absorption coefficient
\end{description}

In our model, the gravity $g$ is computed from the potential theory that
takes into account the ellipticity of the considered planet and the
centrifugal force produced by its rotation. Given the planet's size, mass
and ellipticity and rotational period, our code can calculate gravity at any
latitude on the planet. For Saturn, we obtained 8.95 $m/s^2$ for the
equatorial gravity and 12.06 $m/s^2$ for the polar gravity.

The aerosol opacity $\tau_A$ and aerosol albedo $\varpi_{0A}$ are given as free parameters in aerosol-containing layers.    

As a consequence, the total optical thickness and effective single
scattering albedo of a layer are generally obtained as : 

\begin{equation} 
\tau_{total} = \tau_R + \tau_{A}
\end{equation}

\begin{equation} 
\varpi_{0eff} = \displaystyle{\frac{\tau_R \cdot \varpi_{0R} + \tau_A \cdot \varpi_{0A}}{\tau_{total}}}.
\end{equation}
\vspace{1cm}

\centerline{\bf\Large C: Transmission function of the AOTF} 
\vspace{5mm}

The transmission function $T(\nu)$ of the AOTF at the wavenumber $\nu$
$[cm^{-1}]$ is formulated in the following way.

\begin{equation} 
T(\nu) = \left( \frac{\sin X}{X} \right)^2
\end{equation}

where,

\begin{equation} 
X \equiv \frac{2.784 \cdot (\nu - \nu_0)}{44.0}
\end{equation}

 Here, $\nu_0$ is the wavenumber in $[cm^{-1}]$ at which the AOTF is
 tuned to have the maximum transmission. The conversion between
 $\lambda [nm]$ and $\nu [cm^{-1}]$ simply follows,

\begin{equation} 
\nu [cm^{-1}] = \frac{1}{\lambda [nm] \times 10^{-7}}
\end{equation}
\vspace{1cm}

\centerline{\large\bf Acknowledgments}
\vspace{5mm}

We thank the entire MSSC staff for their help in our observations. In
particular, we express our sincere gratitude to Dr. Lewis Roberts for
his great assistance in the arrangement of our observations. The US Air
Force provided the telescope time, on-site support and 80\% of research
funds for this AFOSR and NSF jointly sponsored research under grant
number NSF AST-0123443. N.J.C. acknowledges support from the Tombaugh
Scholars Program.  
\vspace{1cm}

\centerline{\Huge\bf References}
\everypar={\hangafter=1 \hangindent=.5in}
\vspace{5mm}

\noindent Acarreta, J. R., and A. S\'anchez-Lavega 1999.
	{Vertical cloud structure in Saturn's 1990 equatorial storm.}
	{\it Icarus} {\bf 137}, 24--33.
\vspace{2mm}

\noindent Atreya, S. K., M. H. Wong, T. C. Owen,
	P. R. Mahaffy, H. B. Niemann, I. de Pater, P. Drossart, and
	T. Encrenaz 1999. {A comparison of atmospheres of Jupiter and
	Saturn: deep atmospheric composition, cloud structure, vertical
	mixing, and origin.} {\it Planet. Space. Sci.} {\bf 47},
	1243--1262.   
\vspace{2mm}

\noindent Atreya, S. K., P. R. Mahaffy, H. B. Niemann,
	M. H. Wong, and T. C. Owen 2003. {Composition and origin of the 
	atmosphere of Jupiter - an update, and implications for the
	extrasolar giant planets.} {\it Planet. Space. Sci.} {\bf 51},
	105--112. 
\vspace{2mm}

\noindent Banfield, D., P. J. Gierasch, S. W. Squyres,
	P. D. Nicholson, B. J. Conrath, and K. Matthews 1996. {2$\mu$m
	Spectroscopy of Jovian stratospheric aerosols - Scattering
	opacities, vertical distributions, and wind speeds.} {\it
	Icarus} {\bf 121}, 389--410.  
\vspace{2mm}

\noindent Benassi, M., R. D. M. Garcia, A. H. Karp, and
	C. E. Siewert 1984. {A high-order spherical harmonics solution to
	the standard problem in radiative transfer.} {\it Astrophys. J.}
	{\bf 280}, 853--864.
\vspace{2mm}

\noindent B\'ezard, B., J. P. Baluteau, and A. Marten
	1983. {Study of the deep cloud structure in the equatorial
	region of Jupiter from Voyager infrared and visible data.} {\it
	Icarus} {\bf 54}, 434--455.
\vspace{2mm}
 
\noindent Bohren, C. F. and D. R. Hoffman 1983. {\it
	Absorption and Scattering of Light by Small Particles}, John
	Wiley \& Sons, New York.
\vspace{2mm}

\noindent Bevington, P. R, and D. K. Robinson 1992. {\it
	Data Reduction and Error Analysis for the Physical Sciences},
	McGraw-Hill, Boston.
\vspace{2mm}

\noindent Carlson, B. E., A. A. Lacis, and W. B. Rossow
	1993. {Tropospheric gas composition and cloud structure of the
	Jovian north equatorial belt.}
	{\it J. Geophys. Res.} {\bf 98, E3}, 5251--5290.  
\vspace{2mm}

\noindent Carlson, B. E., A. A. Lacis, and W. B. Rossow
	1994. {Belt-zone variations in the Jovian cloud structure.}
	{\it J. Geophys. Res.} {\bf 99, E7}, 14623--14658.  
\vspace{2mm}

\noindent Chandrasekhar, S. 1960. {\it Radiative Transfer},
	Dover, New York. 
\vspace{2mm}

\noindent Chanover, N. J., C. M. Anderson, C. P. McKay,
	P. Rannou, D. A. Glenar, J. J. Hillman, and W. E. Blass 2003. 
	{Probing Titan's lower atmosphere with acousto-optic tuning.}
	{\it Icarus} {\bf 163}, 150--163.  
\vspace{2mm}

\noindent Conrath, B. J., D. Gautier, R. A. Hanel, and
	J. S. Hornstein 1984. {The helium abundance of Saturn from
	Voyager measurements} {\it Astrophys. J.} {\bf 282}, 807--815. 
\vspace{2mm}

\noindent Conrath, B. J., and D. Gautier 2000. {Saturn helium
	abundance: A reanalysis of Voyager measurement.} {\it Icarus}
	 {\bf 144}, 124--134.  
\vspace{2mm}

\noindent Courtin, R., D. Gautier, A. Marten, B. B\'ezard,
	and R. Hanel 1984. {The composition of Saturn's atmosphere at 
	northern temperate latitudes from Voyager IRIS spectra: $NH_3$,
	$PH_3$, $C_2H_2$, $C_2H_6$, $CH_3D$, $CH_4$, and the Saturnian
	D/H isotopic ratio.} {\it Astrophys. J.} {\bf 287}, 899--916. 
\vspace{2mm}

\noindent Cox, A. N. 2000. {\it Allen's Astrophysical Quantities},
	Springer-Verlag, New York.  
\vspace{2mm}

\noindent Dollfus, A. 1996. {Saturn's rings: Optical
	reflectance polarimetry.} {\it Icarus} {\bf 124}, 237--261. 
\vspace{2mm}

\noindent Dulgach, J. M., A. V. Morozhenko,
	A. P. Vid'machenko, and E. G. Yanovitskij 1983. {Investigations
	of the optical properties of Saturn's atmosphere carried out at
	the main astronomical observatory of the Ukrainian Academy of
	Sciences.} {\it Icarus} {\bf 54}, 319--336.  
\vspace{2mm}

\noindent Esposito, L. W., J. N. Cuzzi, J. B. Holberg,
	E. A. Marouf, G. L. Tyler, and C. C. Porco 1984. {Saturn's rings:
	Structure, dynamics and particle properties.} In {\it Saturn}
	(T. Gehrels and M. S. Matthews, Eds.) pp. 463--545. University of
	Arizona Press, Tucson.  
\vspace{2mm}
  
\noindent Georgiev, G., D.A. Glenar, and J. J. Hillman 2002. {Spectral
characterization of acousto-optic filters used in imaging spectroscopy.}
{\it Appl. Optics} {\bf 41}, 209--217.  
\vspace{2mm}

\noindent Glenar, D. A., J. J. Hillman, B. Saif, and
	J. Bergstralh 1994. {Acousto-optic imaging spectropolarimetry
	for remote sensing.} {\it Appl. Optics} {\bf 33}, No.31,
	7412--7424. 
\vspace{2mm}

\noindent Glenar, D.A., J. J. Hillman, M. Lelouarn,
	R. Q. Fugate, and J. D. Drummond 1997. {Multispectral imagery of 
	Jupiter and Saturn using adaptive optics and acousto-optic
	tuning.} {\it Publ. Astron. Soc. Pac.} {\bf 109}, 326--337. 
\vspace{2mm}

\noindent Hanel, R., B. Conrath, F. M. Flasar, V. Kunde,
	W. Maguire, J. Pearl, J. Pirraglia, R. Samuleson, L. Herath,
	M. Allison, D. Cruikshank, D. Gautier, P. Gierasch, L. Horn,
	R. Koppany, and C. Ponnamperuma 1981. {Infrared
	observations of the Saturnian system from Voyager 1.} {\it
	Science} {\bf 212}, 192--200. 
\vspace{2mm}

\noindent Hansen, J. E. 1969. {Radiative transfer by doubling
	very thin layers.} {\it Astrophys. J.} {\bf 155}, 565--573. 
\vspace{2mm}

\noindent Hansen, J.E. and L.D.Travis 1974. {Light scattering in
	planetary atmospheres.} {\it Space Sci. Rev.} {\bf 16},
	527--610.   
\vspace{2mm}

\noindent Hardorp, J. 1980. {The Sun among the stars
	III. Energy distributions of 16 northern G-type stars and the
	solar flux calibration.} {\it Astron. Astrophys.} {\bf 91},
	221--232.  
\vspace{2mm}

\noindent Karkoschka, E. and M. G. Tomasko 1992. {Saturn's
	upper troposphere 1986--1989.} {\it Icarus} {\bf 97}, 161--181. 
\vspace{2mm}

\noindent Karkoschka, E. and M. G. Tomasko 1993. {Saturn's
	upper atmospheric hazes observed by the Hubble Space Telescope.}
	{\it Icarus} {\bf 106}, 428--441.  
\vspace{2mm}

\noindent Karkoschka, E. 1994. {Spectrophotometry of the
	Jovian planets and Titan at 300- to 1000-nm wavelength: The
	methane spectrum.} {\it Icarus} {\bf 111}, 174--192. 
\vspace{2mm}

\noindent Lindal, G. F., D. N. Sweetnam, and V. R. Eshleman
	1985. {The atmosphere of Saturn: An analysis of the Voyager
	radio occultation measurements.} {\it Astron. J.} {\bf 90},
	1136--1146.  
\vspace{2mm}

\noindent Marten, A., D. Rouan, J. P. Baluteau, D. Gautier,
	B. J. Conrath, R. A. Hanel, V. Kunde, R. Samuelson, A. Chedin,
	and N. Scott 1981. {Study of the ammonia ice cloud layer in the
	equatorial region of Jupiter from the infrared interferometric
	experiment on Voyager.} {\it Icarus} {\bf 46}, 233--248. 
\vspace{2mm}

\noindent O'Brien, J.J., and H. Cao 2002. {Absorption spectra and
absorption coefficients for methane in the 750--940 nm region obtained
by intracavity laser spectroscopy.} {\it
J. Quant. Spectrosc. Radiat. Transf.} {\bf 75}, 323--350.  
\vspace{2mm}

\noindent Ortiz, J. L., F. Moreno, and A. Molina 1995. {Saturn 
	1991-1993: Reflectivities and limb-darkening coefficients at
	methane band and nearby continua -- Temporal changes.} {\it
	Icarus} {\bf 117}, 328--344. 
\vspace{2mm}

\noindent Ortiz, J. L., F. Moreno, and A. Molina 1996. {Saturn 
	1991-1993: Clouds and hazes.} {\it Icarus} {\bf 119}, 53--66.
\vspace{2mm}
 
\noindent Potter, J. F. 1970. {The delta function approximation
	in radiative transfer theory.} {\it J. Atmos. Sci.} {\bf
	27}, 943--949.  
\vspace{2mm}

\noindent Sanchez-Lavega, A., J. Lecacheux, F. Colas, and P. Laques
	1994. {Photometry of Saturn's 1990 equatorial disturbance} {\it
	Icarus} {\bf 108}, 158--168. 
\vspace{2mm}

\noindent Santer, R., and A. Dollfus 1981. {Optical
	reflectance polarimetry of Saturn's globe and rings : IV. Aerosol
	in the upper atmosphere of Saturn.} {\it Icarus} {\bf 48},
	496--518. 
\vspace{2mm}

\noindent Singh, K., and J. J. O'Brien 1995. {Laboratory measurements of
	absorption coefficients for the 727 nm band of methane at 77 K
	and comparison with results derived from spectra of the giant
	planets.} {\it J. Quant. Spectrosc. Radiat. Transf.} {\bf 54},
	607--619.  
\vspace{2mm}

\noindent Sromovsky, L. A., and P. M. Fry 2002. {Jupiter's 
	cloud structure as constrained by Galileo probe and HST
	observations.} {\it Icarus} {\bf 157}, 373--400.  
\vspace{2mm}

\noindent Stam, D. M., D. Banfield, P. J. Gierasch, P. D. Nicholson, and
K. Matthews. 2001. {Near-IR spectrophotometry of Saturnian aerosols --
meridional and vertical distribution.} {\it Icarus} {\bf 152}, 407--422. 
\vspace{2mm}

\noindent Stammes, P., J. F. de Haan, and J. W. Hovenier
	1989. {The polarized internal radiation field of a planetary
	atmosphere.} {\it Astron. Astrophys.} {\bf 225}, 239--259. 
\vspace{2mm}

\noindent Stamnes, K., S. C. Tsay, W. Wiscombe, and
	K. Jayaweera 1988. {Numerically stable algorithm for
	discrete-ordinate-method radiative transfer in multiple
	scattering and emitting layered media.} {\it Appl. Optics} {\bf
	27}, 2502--2509. 
\vspace{2mm}

\noindent Thomas, G.E., and K. Stamnes 1999. {\it Radiative
	Transfer in the Atmosphere and Ocean}, Cambridge University
	Press, Cambridge, U.K. 
\vspace{2mm}

\noindent Tomasko, M.G., R. S. McMillan, L. R. Doose,
	N. D. Castillo, J. P. Dilley 1980. {Photometry of Saturn at
	large phase angles.} {\it J. Geophys. Res.} {\bf 85, A11},
	5891--5903. 
\vspace{2mm}

\noindent Tomasko, M.G. and L. R. Doose 1984. {Polarimetry 
	and Photometry of Saturn from Pioneer 11 : Observations and
	constraints on the distribution and properties of cloud and
	aerosol particles.} {\it Icarus} {\bf 58}, 1--34.
\vspace{2mm}

\noindent Weidenschilling, S. J., and J. S. Lewis
	1973. {Atmospheric and cloud structures of the Jovian planets.}
	{\it Icarus} {\bf 20}, 465--476. 
\vspace{2mm}

\noindent West, R. A., M. G. Tomasko, B. A. Smith, M. P. Wijesinghe,
	L. R. Doose, H. J. Reitsema, and S. M. Larson 1982. {Spatially
	resolved methane band photometry of Saturn I, Absolute
	reflectivity and center-limb variations in the 6190-, 7250-, and
	8900-\AA bands.} {\it Icarus} {\bf 51}, 51--64.  
\vspace{2mm}

\noindent West, R.A. 1983. {Spatially resolved methane band
	photometry of Saturn II, Cloud structure models at four
	latitudes.} {\it Icarus} {\bf 53}, 301--309. 
\vspace{2mm}

\noindent West, R.A., M. Sato, A. L. Hart, C. W. Hord,
	K. E. Simmons, L. W. Esposito, D. L. Coffeen, R. B. Pomphrey
	1983. {Photometry and polarimetry of Saturn at 2640 and 7500
	\AA.} {\it J. Geophys. Res.} {\bf 88, A11}, 8679--8697.

\end{document}